\documentclass[10pt]{iopart}

\usepackage{ascmac}
\usepackage{cite}
\usepackage{comment}
\usepackage{enumerate}

\usepackage{graphicx}
\newcommand{\mathsym}[1]{{}}

\newcommand{\si}{$\rm{i}$}
\newcommand{\sii}{$\rm{i\hspace{-.1em}i\hspace{-.1em}}$}
\newcommand{\siii}{$\rm{i\hspace{-.1em}i\hspace{-.1em}i}$}
\newcommand{\siv}{$\rm{i\hspace{-.1em}v}$}
\newcommand{\sv}{$\rm{\hspace{-.1em}v}$}

\newcommand{\1}{\mbox{1}\hspace{-0.25em}\mbox{l}}
\newcommand{\dsp}{\displaystyle}
\usepackage{ulem}
\usepackage{color}

\usepackage{caption}

\usepackage{iopams}

\begin{document}

\title{
Exact Bethe quantum numbers of the massive XXZ chain 
in the two down-spin sector} 
{For the solutions the spin-1/2 chain 
 }

\author{Takashi Imoto${}^{1}$, Tetsuo Deguchi${}^{2}$}
\address{${}^{1}$Research Center for Emerging Computing Technologies, National Institute of Advanced Industrial Science and Technology (AIST), 1-1-1 Umezono, Tsukuba, Ibaraki 305-8568, Japan.\\
${}^{2}$Department of Physics, Ochanomizu University, 2-1-1 Ohtsuka, Bunkyo-ku, Tokyo 112-8610, Japan}

\ead{deguchi@phys.ocha.ac.jp; takashi.imoto@aist.go.jp}
\vspace{10pt}
\begin{indented}
\item[]\today
\end{indented}

\begin{abstract}
Every solution of the Bethe ansatz equations (BAE) is characterized by a set of quantum numbers called the Bethe quantum numbers, which are fundamental for evaluating it numerically. 
We rigorously derive the Bethe quantum numbers for the real solutions of the spin-1/2 massive XXZ spin chain in the two down-spin sector, assuming the existence of solutions to some form of BAE.
In the sector 
the quantum numbers $J_1$ and $J_2$
were derived for complex solutions, but not for real solutions. 
%
%
We show the exact results in the sector as follows.
%
(\si) When two Bethe quantum numbers are different, 
i.e. for $J_1 \ne J_2$, 
we introduce a graphical method, which we call a contour method, for deriving the solution of BAE to a given set of Bethe quantum numbers. 
By the method, we can readily show the existence and the uniqueness of the solution.
(\sii) When two Bethe quantum numbers are equal, 
i.e. for $J_1 = J_2$,
we derive the criteria for the collapse of two-strings and the emergence of an extra two-string by an analytic method.
(\siii) We obtain the number of real  
solutions, which depends on the site number $N$ and the XXZ anisotropy parameter $\zeta$. 
(\siv) We derive all infinite-valued solutions of BAE for the XXX spin chain in the two down-spin sector through the XXX limit. 
(\sv) We explicitly show the completeness of the Bethe ansatz in terms of the Bethe quantum numbers.

%
\end{abstract}

\vspace{2pc}

\section{Introduction}

The Heisenberg spin chain (also called the spin-1/2 XXX spin chain) and the spin-1/2 XXZ spin chain are fundamental integrable models in both condensed matter physics and mathematical physics.
%
The Bethe ansatz is a powerful method for solving quantum integrable models\cite{Bethe, Baxter1982, FT}. 
In the method, the Bethe ansatz equations play a central role. From a solution of the Bethe ansatz equations, we derive the corresponding eigenvalue and eigenvector of the quantum Hamiltonian.
However, it is not trivial to obtain all the solutions of the Bethe ansatz equations numerically even in a restricted sector other than the zero and one down-spin sectors. 

The Hamiltonian of the spin-1/2 anisotropic quantum Heisenberg spin chain, i.e., the spin-1/2 XXZ spin chain under the periodic boundary conditions is given by
\begin{eqnarray}
H_{XXZ}=\frac{1}{4}\sum_{j=1}^{N}\biggl(\sigma_{j}^{x}\sigma_{j+1}^{x}+\sigma_{j}^{y}\sigma_{j+1}^{y}+\Delta\bigl(\sigma_{j}^{z}\sigma_{j+1}^{z}-\1\bigr)\biggr)
\end{eqnarray}
where $\sigma_{j}^{a}(a=x,y,z)$ are the Pauli matrices acting on the $j$th site of the chain, $\Delta$ denotes the XXZ anisotropy parameter, and $N$ the site number. 
It is known that when $|\Delta|>1$ the energy spectrum has a gap at the ground state, while when $|\Delta|\leq1$ it is gapless. In particular, when $\Delta=1$, this model is called by the XXX spin chain.

In the $M$ down-spin sector with rapidities $\lambda_{1}, \lambda_2, \cdots,\lambda_{M}$, 
the Bethe ansatz equations (BAE) for the spin-1/2 XXZ spin chain are given by
\begin{eqnarray}
\biggl(\frac{\phi(\lambda_{j}+i\zeta/2)}{\phi(\lambda_{j}-i\zeta/2)}\biggr)^{N}=\prod_{k\neq j,k=1}^{M}\frac{\phi(\lambda_{j}-\lambda_{k}+i\zeta)}{\phi(\lambda_{j}-\lambda_{k}-i\zeta)}\ \ (j=1,2,\cdots,M).\label{eq:BAE_mul}
\end{eqnarray}
If the anisotropy parameter $\Delta$ is equal to 1: $\Delta = 1$ (i.e., the XXX spin chain), we define $\phi$ and $\zeta$ as $\phi(\lambda)=\lambda$ and $\zeta=1$. If $\Delta>1$(i.e., the massive XXZ spin chain), we define $\phi$ and $\zeta$ as $\phi(\lambda)=\sin(\lambda)$ and $\Delta=\cosh(\zeta)$. 
If $-1<\Delta<1$(i.e., the  massless XXZ spin chain), we define $\phi$ and $\zeta$ as $\phi(\lambda)=\sinh(\lambda)$ and $\Delta=\cos(\zeta)$.   
In this paper, we also refer to $\zeta$ as the anisotropic parameter.

In order to define 
the Bethe quantum numbers, we take the logarithm for both hand sides of BAE (\ref{eq:BAE_mul}).  
For $\Delta>1$ we derive
\begin{eqnarray}
2\tan^{-1}\biggl(\frac{\tan(\lambda_{i})}{\tanh(\zeta/2)}\biggr)=\frac{2\pi}{N}J_{i}+\frac{1}{N}\sum_{k=1}^{M}2\tan^{-1}\biggl(\frac{\tan(\lambda_{i}-\lambda_{k})}{\tanh(\zeta)}\biggr),\\
J_{i}\equiv\frac{1}{2}(N-M+1)\ (\mbox{mod}\ 1)\ \ \mbox{for}\ i=1,2,\cdots,M. \label{eq:mod}
\end{eqnarray}
We call $J_{i}$ the Bethe quantum numbers. 
In eq. (\ref{eq:mod}) $J_i$ are half-integers if $N-M$ is even, while integers if $N-M$ is odd. 
The Bethe quantum numbers specify the selected sheet in the Riemann surface of the logarithmic function, on which the solution of BAE is defined. 
Each function appearing in the logarithmic form of BAE is defined on it via analytic continuation, and it has to be consistent with each other.
Therefore, the Bethe quantum numbers are not given by arbitrary integers or half-integers. 
If a set of the Bethe quantum numbers for a solution of BAE is specified, we can derive it by numerical methods\cite{HC}. 
There is a set of conjectures on the range of Bethe quantum numbers called the string hypothesis \cite{TK1,TK}, which will be explained in section 2.   
However, it is not trivial to select valid sets of Bethe quantum numbers. In fact, some complex solutions may collapse into real solutions \cite{EKS, DG1}, so that the string hypothesis is violated, as discussed in section 2. We remark that the completeness of the Bethe ansatz for the spin-1/2 XXX spin chain has been studied \cite{KE,Kiri1,Kiri2,Baxter,LYSA,T1,MTV}.


There are several physical motivations for the study of the Bethe quantum numbers in the two down-spin sector (i.e., for $M=2$) 
for the spin-1/2 XXZ spin chain in the massive regime. 
(\si) If all sets of the Bethe quantum numbers in the sector are obtained,  we can express any given state as a linear combination of the Bethe eigenvectors in the sector. Making use of the Bethe quantum numbers we can evaluate numerical solutions to BAE, in particular, in the logarithmic form. 
(\sii) 
We can perform exact quantum dynamics in a finite XXZ spin chain 
with an arbitrary site number $N$ in the two down-spin sector ($M=2$) by expressing an initial quantum state as a linear combination of Bethe eigenvectors.


In this paper, we analytically show the number of real solutions by deriving all Bethe quantum numbers in the two down-spin sector for the massive regime of the spin-1/2 XXZ spin chain with an arbitrary site number $N$. In the sector 
we demonstrate exact results as follows. (\si) When two Bethe quantum numbers $J_1$ and $J_2$ are different, we introduce a graphical method for deriving the solution of BAE denoted by $\lambda_1$ and $\lambda_2$ which corresponds to a given set of Bethe quantum numbers $J_1$ and $J_2$, respectively. 
(\sii) When two Bethe quantum numbers are equal, we derive the criteria for the collapse of two-strings and the emergence of an extra two-string by an analytic method.
(\siii) We thus derive all the Bethe quantum numbers for the one-string (i.e., real) solutions; 
(\siv) The number of one-string (i.e., real) solutions depends on the site number $N$ and the XXZ-anisotropy parameter $\zeta$; 
(\sv) We derive an infinite-valued solution of BAE in the XXX limit. 
Throughout the paper we assume that the site number $N$ is even, although the method is applicable to the odd $N$ case.

\begin{figure}[ht]
  \includegraphics[clip,width=15.0cm]{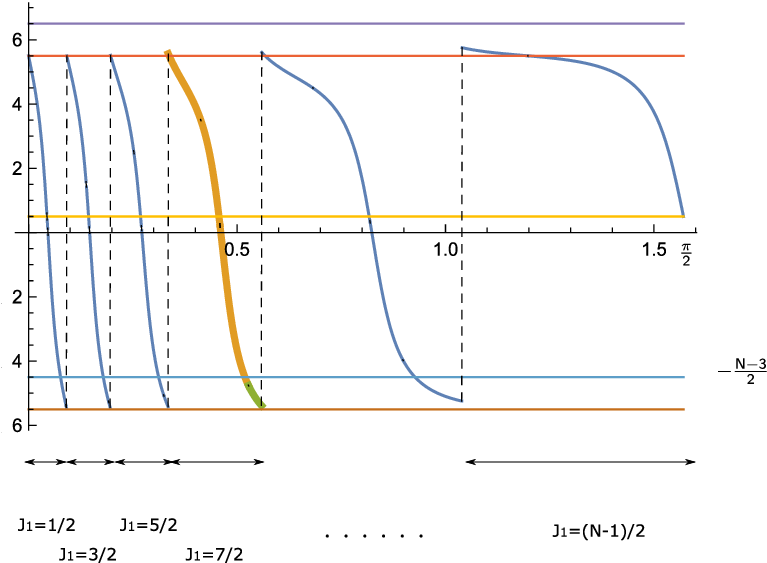}
  \caption{
  Graphs of height function $h(\zeta, \mu_{1})$ with fixed $J_1$'s versus rapidity $\mu_{1}$. Horizontal axis (i.e., $x$-axis) denotes rapidity $\mu_{1}$, 
while the vertical axis (i.e., $y$- axis) denotes values of $h(\zeta, \mu_{1})$ for $\zeta=0.7$ and $N=12$. 
The orange bold line is part of the graph corresponding to $J_{1}=7/2$ on the domain of definition in eq. (\ref{eq:domain1}), while the green bold line is part of the graph with $J_{1}=7/2$ on the domain of definition in eq. (\ref{eq:domain2}). Thus, the whole graph of height function $h(\zeta, \mu_{1})$ with $J_{1}=7/2$ for $-\pi/2 < \mu_1 - \mu_2 < \pi$ is given by a simply connected contour from the top at $J_2= 11/2$ to the bottom at $J_2=-11/2$. Here the range of $\mu_1$ corresponding to $J_1=7/2$ is restricted in the interval from about 0.32 to about 0.57. Hence, the intersection of a horizontal line $y=J_2$ and the graph of $J_1$ leads to the solution of BAE for quantum numbers $J_1$ and $J_2$. For instance, the graph of $y=J_2=1/2$, depicted by the yellow horizontal line,  intersects with the orange bold line at $\mu_1 \approx 0.4$.  
We remark that the largest quantum number $J_1$ is given by $(N-1)/2$ as we shall analytically  show in section 5.}
  \label{fig:h(zeta)_one}
\end{figure}

Let us illustrate the graphical method for case (i) of the last paragraph. Suppose that  $\mu_1$ denotes a continuous variable called rapidity.  We introduce a function of rapidity $\mu_1$ with fixed quantum number $J_1$ denoted by $h(J_1; \mu_1)$, which we call a height function, such that it gives the Bethe quantum number $J_2$ when rapidity $\mu_1$ is equal to $\lambda_1$ in the solution ($\lambda_1$,$\lambda_2$) of BAE for quantum numbers $J_1$ and $J_2$: $J_2=h(J_1; \lambda_1)$.
We show that the graph of $h(J_1; \mu_1)$ versus $\mu_1$ with fixed $J_1$ is expressed as a simply connected contour in the $xy$ plane where the $x$- and $y$- axis denote rapidity $\mu_1$ and the value of $h(J_1; \mu_1)$, respectively, as shown in Figure \ref{fig:h(zeta)_one}. Thus, an intersection between the contour of $J_1$ and the graph $y=J_2$ determines the value $\lambda_1$ for the solution to BAE of quantum numbers $J_1$ and $J_2$. 
Here we remark that another one, $\lambda_2$, is derived from $\lambda_1$
through the BAE, which we shall formulate shortly.  

%

We now explicitly explain the method in this manuscript.
In terms of Gauss' symbol
we define the Bethe quantum numbers $J_1$ and $J_2$ 
for the Bethe-ansatz equations
in the two down-spin sector of the spin-$1/2$ massive XXZ spin chain with an arbitrary even number $N$ of sites \cite{TK}.
\begin{eqnarray}
 2\tan^{-1}&\biggl(\frac{\tan\lambda_{1}}{\tanh{\zeta/2}}\biggr)=\frac{2\pi}{N}J_{1}\nonumber\\
&+\frac{2}{N}\tan^{-1}\biggl(\frac{\tan(\lambda_{1}-\lambda_{2})}{\tanh\zeta}\biggr)+\frac{2\pi}{N}\biggl[\frac{2(\lambda_{1}-\lambda_{2})+\pi}{2\pi}\biggr]_{Gauss},\label{eq:BAE11}
\\
2\tan^{-1}&\biggl(\frac{\tan\lambda_{2}}{\tanh{\zeta/2}}\biggr)=\frac{2\pi}{N}J_{2}\nonumber\\
&+\frac{2}{N}\tan^{-1}\biggl(\frac{\tan(\lambda_{2}-\lambda_{1})}{\tanh\zeta}\biggr)+\frac{2\pi}{N}\biggl[\frac{2(\lambda_{2}-\lambda_{1})+\pi}{2\pi}\biggr]_{Gauss}\label{eq:BAE22}.
\end{eqnarray}
The symbol $[x]_{Gauss}$ denotes the greatest integer that is not larger than $x$.
We shall show that it plays a role in the graphical method.

We derive all sets of Bethe quantum numbers $J_1$ and $J_2$ for real solutions in the two down-spin sector by two different methods depending on whether $J_1$ and $J_2$ are equal or not. 
We call it case I when $J_1$ and $J_2$ are different ($J_1 \ne J_2$) and case II when they are equal  ($J_1=J_2$).  
We show case I in the part I and case II in the part II of the present manuscript. 
%


In the part I we introduce another continuous variable  $\mu_2$ in addition to rapidity $\mu_1$. We shall show that 
any real solution $\lambda_1$ and $\lambda_2$ of the BAE 
(\ref{eq:BAE11}) and (\ref{eq:BAE22}) corresponds to a special case of $\mu_1$ and $\mu_2$, respectively.
Let us now assume that rapidities $\mu_1$ and $\mu_2$ satisfy only the first equation (\ref{eq:BAE11}) among eqs. (\ref{eq:BAE11}) and (\ref{eq:BAE22}), while they do not necessarily satisfy the second equation (\ref{eq:BAE22}). They thus satisfy the following: 
\begin{eqnarray}
& \tan\mu_{1}/\tanh{\zeta/2} & \nonumber \\
= &  \tan\left( 
\frac{\pi}{N}J_{1} + \frac{1}{N}\tan^{-1}\biggl(\frac{\tan(\mu_{1}-\mu_{2})}{\tanh\zeta}\biggr)+\frac{\pi}{N}\biggl[\frac{2(\mu_{1}-\mu_{2})+\pi}{2\pi}\biggr]_{Gauss} 
\right) &, \nonumber \\
\label{eq:BAE11-tan}
\end{eqnarray}
Furthermore, 
we first consider the case when the first rapidity of a solution $\lambda_{1}$ is positive, i.e., only the case of $\mu_1 >0$. 
\footnote{If both the signs of $\lambda_{1}$ and $\lambda_{2}$ are changed to other signs, $\lambda_{1}$ and $\lambda_{2}$ correspond to Bethe quantum numbers of opposite signs. In other words, if the solution of Bethe ansatz equation is changed as  $(\lambda_{1},\lambda_{2})\rightarrow(-\lambda_{1},-\lambda_{2})$, the Bethe quantum numbers convert to $(-J_{1}, -J_{2})$. Thus, we restrict our analysis to the case where $\lambda_{1}$ is positive.}
The case of $\mu_{1} < 0$ 
will be discussed in \ref{sec:negative_lambda1}. 
Thus, the range of $\mu_1-\mu_2$ is given by 
\begin{equation}
    -\pi/2 < \mu_1-\mu_2 < \pi
\end{equation}
since  $0 < \mu_1 < \pi/2$ and $-\pi/2 < \mu_2 < \pi/2$. 
When $\mu_1$ is positive, we consider 
two regions of $\mu_1-\mu_2$ as follows.  
\begin{equation}
-\pi/2 < \mu_1 - \mu_2 < \pi/2 \quad \mbox{and} \quad 
\pi/2 < \mu_1 - \mu_2 < \pi .  \label{eq:regions}
\end{equation}
We therefore consider two domains of definition for rapidity $\mu_1$ as follows. 
\par \noindent 
(\si) For $\mu_{1}-\mu_{2}<\frac{\pi}{2}$ 
\begin{eqnarray}
\tanh(\zeta/2)\tan\biggl(\frac{\pi}{N}(J_{1}-\frac{1}{2})\biggr)<\tan&(\mu_{1})<\tanh(\zeta/2)\tan\biggl(\frac{\pi}{N}(J_{1}+\frac{1}{2})\biggr)\nonumber\\
\label{eq:domain1}
 \end{eqnarray}
(\sii) For $\mu_{1}-\mu_{2}>\frac{\pi}{2}$
\begin{eqnarray}
\tanh(\zeta/2)\tan\biggl(\frac{\pi}{N}(J_{1}+\frac{1}{2})\biggr)<\tan&(\mu_{1})<\tanh(\zeta/2)\tan\biggl(\frac{\pi}{N}(J_{1}+\frac{3}{2})\biggr)\nonumber\\
\label{eq:domain2}
\end{eqnarray}

Let us now regard $\mu_2$ as a function of $\mu_1$. Here we recall that $\mu_1$ and $\mu_2$ satisfy only the first equation (\ref{eq:BAE11}) of BAE, i.e. eq. (\ref{eq:BAE11-tan}).
%
We denote by $\mu_2(\mu_1)$ rapidity $\mu_2$ as a function of $\mu_1$. 
We define function $h(J_1; \zeta, \mu_{1})$, 
which we have called the height function, by the following:   
\begin{eqnarray}
h(J_1; \zeta, \mu_{1})\equiv&&\frac{N}{\pi}\tan^{-1}\biggl(\frac{\tan\mu_{2}(\mu_1) }{\tanh\zeta/2}\biggr)\nonumber\\
&&-\frac{1}{\pi}\tan^{-1}\biggl(\frac{\tan(\mu_{2}(\mu_1)-\mu_{1})}{\tanh\zeta}\biggr)
-\biggl[\frac{2(\mu_{2}(\mu_1) -\mu_{1})+\pi}{2\pi}\biggr]_{Gauss}.\label{eq:def_search_func}
\end{eqnarray}
where $J_{1}$ satisfies the condition (\ref{eq:domain1}) or (\ref{eq:domain2}). 
We can show that if we solve the equation: $J_{2}=h(J_1; \zeta, \mu_{1})$ with respect to $\mu_1$, we obtain the solution of the Bethe ansatz equations (\ref{eq:BAE11}) and (\ref{eq:BAE22}) for the Bethe quantum numbers  $J_{1}$ and $J_{2}$. 
Thus, the equation with respect to $\mu_1$: $J_{2}=h(J_1; \zeta, \mu_{1})$ corresponds to the second equation (\ref{eq:BAE22}) of the Bethe ansatz equations. 


In the part II we introduce another function, which we call the counting function.  
\footnote{
If the Bethe quantum numbers are equal: $J_{1}=J_{2}$ 
it is not straightforward to use the height function,  
since the solutions of Bethe ansatz equations include 
the trivial solution such as $\lambda_{1}=\lambda_{2}$.}  
Let us assume that real solutions of the Bethe ansatz equations satisfy the following form:  
\begin{eqnarray}
\lambda_{1}&=x-\frac{1}{2}\gamma\zeta \\
\lambda_{2}&=x+\frac{1}{2}\gamma\zeta.
\end{eqnarray}
We define a counting function $W(x, \gamma, \zeta)$ by 
\begin{eqnarray}
2\pi W(x,\gamma,\zeta)&&\equiv \tan^{-1}\biggl(\frac{\tan(x-\frac{1}{2}\gamma\zeta)}{\tanh(\frac{\zeta}{2})}\biggr)\nonumber\\
&&+\tan^{-1}\biggl(\frac{\tan(x+\frac{1}{2}\gamma\zeta)}{\tanh(\frac{\zeta}{2})}\biggr)-\frac{2\pi}{N}\biggl[\frac{-2\gamma\zeta+\pi}{2\pi}\biggr]_{\mathrm{Gauss}}.\nonumber\\
\label{eq:definition_of_counting_function}
\end{eqnarray}
We may regard variable $x$ as a function of variable $\gamma$. 
Its expression will be explicitly given in eq.  (\ref{eq:real1}).  
When we solve the equation: $J_{1}=NW(x,\gamma,\zeta)$ where 
the two Bethe quantum numbers $J_1$ and $J_2$ are equal,
we obtain the solution of the Bethe ansatz equation corresponding to the Bethe quantum number $J_{1}$ and $J_{2}$.

For an illustration, the height function is plotted in Figure \ref{fig:h(zeta)_one} for site number $N=12$ and $\zeta=0.7$. 
We note that the Bethe quantum number $J_{1}$ is given by half-integer because the site number $N$ is even.
In the case of $J_{1}=7/2$, we search for a crossing point between a given half-integer $J_2$ and the plot of the height function. 
In Figure \ref{fig:h(zeta)_one}, the bold line corresponds to the plot of the height function for $J_{1}=7/2$.

We remark that the graphical method using contours is useful for deriving numerical solutions of BAE.
In a recursive method it should depend on initial values whether we can finally approach the solution of BAE to a given set of Bethe quantum numbers.

In a previous study, all Bethe quantum numbers pertaining to the two-string solutions of the Bethe ansatz equation were derived\cite{ISD}.
However, this derivation included a numerical calculation for conditions leading to the collapse or emergence of additional two-string solutions.
In this research, we obtain all the Bethe quantum numbers for the one-string solutions bypassing the need for the numerical analysis and various assumptions.
Consequently, we obtain all Bethe quantum numbers for the XXZ spin chain in the two down-spin sector.

The contents of this paper are structured as follows. In Section \ref{sec: prevwork}, we provide a comprehensive review of previous research regarding Bethe quantum numbers with $M = 2$ in the spin-1/2 XXZ spin 
enumerating the solutions of the Bethe ansatz equations for distinct pairs of Bethe quantum numbers ($J_1 \ne _2$), while Sections 6 and 7 address the case of equal Bethe quantum numbers ($J_1 = J_2$).
Here we remark that the part I consists of sections 3, 4 and 5, and the part II sections 6 and 7.
Let us explain the contents of the part I as follows. 
First, section 3 is devoted to expressing rapidity $\mu_{2}$ as a function of another rapidity $\mu_{1}$. In particular, In section 3 
we focus on transforming the height function (\ref{eq:def_search_func}) into a function that depends solely on rapidity $\mu_1$, and it is the motivation for expressing $\mu_2$ as a function of$\mu_1$.
Next, in section 4 we prove the monotonicity of the height function except for singularity point. In addition, we derive both the limit from the right and the limit from the left at the discontinuity point of the height function. 
Section 5 delves into an analysis of the height function's discontinuity and thus we derive all Bethe quantum numbers for $J_1 \ne J_2$.
%
Let us next explain the contents of the part II as follows. Sections 6 and 7 are dedicated to the case where $J_1 = J_2$. In Section 6, we introduce the counting function, and Section 7 focuses on deriving all Bethe quantum numbers that satisfy $J_1 = J_2$, along with conditions for the collapse or emergence of extra two-string solutions.
In Section 8, we derive the conditions under which the Bethe solutions diverge in the XXX limit. Section 9 confirms the completeness of the massive XXZ spin chain in the two down-spin sector and derives all relevant Bethe quantum numbers. Finally, in Section 10, we detail the numerical solution of the Bethe ansatz equations based on the Bethe quantum numbers.

\section{Review on the complex solutions for bound states}\label{sec: prevwork}

We now explain fundamental aspects of complex solutions in the two down-spin sector of the spin-1/2 XXZ spin chain, in particular, exact results obtained in a previous work\cite{ISD}.

The ground-state energy of the antiferromagnetic XXX spin chain was calculated by Hulth\'{e}n, where every rapidity is real in the Bethe ansatz solution \cite{Hul,Or,Wa,YY}. 
On the other hand, in excited states at finite temperatures, complex solutions appear in the Bethe ansatz equations. In general, it is not known whether one can obtain every set of complex solutions for the Bethe ansatz equations of the XXX or XXZ spin chain. 
Here we remark some earlier important studies on complex solutions of the spin-1/2 XXZ spin chain\cite{Wo,BVV,FM1,FM2}. 

There exists a set of numerical assumptions or conjectures on the forms of complex solutions, which we call the string hypothesis\cite{TK1,TK}. 
By assuming this hypothesis, we can evaluate 
the free energy at finite temperatures at least approximately, and thermal quantities such as specific heats and magnetic susceptibilities\cite{TY,YT,Sch2,Sch1,Z1,ZNP}. 
Here we remark that there are combinatorial approaches for classifying solutions of the Bethe ansatz equations\cite{KKR, KR, KS1, KS2, DG2, GD2}.
It is believed that physical quantities in the bulk order evaluated by the string hypothesis are correct in the thermodynamic limit. 
However, finite-size corrections to them or physical quantities in finite systems can be of poor accuracy. 
In particular, physical quantities during quantum dynamics of finite systems obtained by the string hypothesis can be of low accuracy.

In order to at least approximately evaluate physical quantities of the XXZ chain in the thermodynamic limit,  we introduce the string hypothesis. This hypothesis is composed of two parts. (\si) The solutions of the Bethe ansatz equations assumed by the string hypothesis for the spin-1/2 XXZ spin chain in the massive regime have  the following form, in general:
\begin{eqnarray}
\lambda_{\alpha,j}^{n}=\lambda_{\alpha}^{n}+(n+1-2j)\frac{i\zeta}{2}+O(e^{-dN}),\ j=1,2,\cdots,n\label{eq:stringForm}
\end{eqnarray}
where $d$ is a positive constant and the string center $\lambda_{\alpha}^{n}$ is given by a real number satisfied by $-\pi/2<\lambda_{\alpha}^{n}<\pi/2$. We call a complex solution of the form (\ref{eq:stringForm}) an $n$-string. The set of all solutions in the $M$ down-spin chain is composed of $k-$strings for $k=1,2,\cdots,M$. The total number of $k-$strings is expressed by $M_{k}$ for each $k$. It is clear that we have $M=\sum_{k=1}^{\infty}k M_{k}$. (\sii) The numbers of sets of $k-$strings for $k=1,2,\cdots,M$ are 
determined by assuming the string hypothesis.
The counter examples of this hypothesis are known in the spin 1/2 XXX and massless XXZ spin chains in the two down-spin sector.\cite{GD2, EKS, IP, IKPP}

Let us recall the Bethe ansatz equations 
(\ref{eq:BAE11}) and (\ref{eq:BAE22})
for the spin-$1/2$ massive XXZ spin chain in the two down-spin sector.
In the two down-spin sector, the set of solutions of the Bethe ansatz equation for the massive XXZ spin chain consists of both real and complex solutions. For the massive XXZ spin chain where $\Delta=\cosh\zeta$ and $\delta>-\zeta/2$, we express the complex solution as follows\cite{Vl1}.
\begin{eqnarray}
\lambda_{1}=x+\frac{i}{2}\zeta+i\delta,\ \lambda_{2}=x-\frac{i}{2}\zeta-i\delta\label{eq:string-two-down-spin-complex}
\end{eqnarray}

\begin{figure}[htbp]
\begin{center}
 \includegraphics[clip,width=10cm]{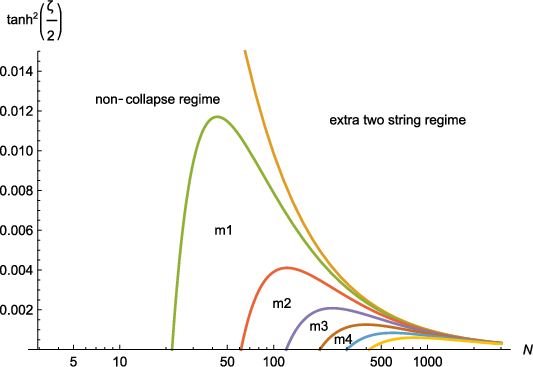}
\end{center}
 \caption{
An extra pair of two-string solutions appears in the area written in extra two-string regime.
Symbol $mk$ such as $m1, m2, \cdots, $ denotes the regime of $k$ missing two string solutions for $k=1,2,\cdots$.
The vertical axis shows the value of $\tanh^{2}(\zeta/2)$ and the horizontal axis the number of sites $N$ on a logarithmic scale. The XXZ anisotropy $\Delta$ is given by $\Delta = \cosh\zeta$.
These regimes are indicated by the eqs.(\ref{eq:condition_of_extra-two-string_form_complex_solution}) and (\ref{eq:condition_of_N_collapse_form_complex_solution}).
}
 \label{fig:extra-sol} 
\end{figure}

Some of two-string solutions whose existence is predicted by the string hypothesis become real solutions if the site number $N$ is large, for the spin-$1/2$ XXX chain in the two down-spin sector\cite{EKS, DG1}.
We call it the collapse of two-string solutions to real ones. The critical number $N_{c}$ such that a collapsed two-string solution exists for $N>N_{c}$ is $21.86$ in the XXX spin chain.
The collapse is numerically investigated\cite{FKT}.
The number of the collapsed two-string solutions 
for the XXX spin chain is rigorously obtained\cite{DG1}. 
Furthermore, every Bethe quantum number is known for the XXX spin chain in the two down spin sector\cite{DG1}. On the other hand, if for the massive XXZ spin chain in the two down-spin sector the anisotropic parameter $\zeta$ and the site number $N$ satisfy the inequality:
\begin{eqnarray}
\tanh^{2}(\zeta/2)<\frac{1-(N-1)\tan^{2}(\frac{3\pi}{2N})}{(N-1)-\tan^{2}(\frac{3\pi}{2N})},\label{eq:condition_of_N_collapse_form_complex_solution} 
\end{eqnarray}
the complex solution becomes collapsed two-string solution(i.e., the real solution) as show in the Figure \ref{fig:extra-sol}.
Whereas, if the anisotropy parameter $\zeta$ and the site number $N$ satisfy the inequality:
\begin{eqnarray}
\tanh^{2}(\zeta/2)>\frac{1-(N-1)\tan^{2}(\frac{\pi}{2N})}{(N-1)-\tan^{2}(\frac{\pi}{2N})},\label{eq:condition_of_extra-two-string_form_complex_solution}
\end{eqnarray}
it is known that the additional complex solution appears and a pair of real solutions disappear as shown in Figure \ref{fig:extra-sol}\cite{ISD}.
We call the new complex solutions extra two-string solutions.
The completeness of the XXX and XXZ spin chains in the two down spin sector is known \cite{DG1, KE}.

\section{Expression of rapidity $\mu_{2}$ in terms of $\mu_{1}$} \label{sec: lamb1-lamb2}

We recall that in the part I of the manuscript consisting of sections 3 to 5 we derive the Bethe quantum numbers $J_1$ and $J_2$ for the case where $J_1 \ne J_2$. 

A solution of BAEs (\ref{eq:BAE11}) and (\ref{eq:BAE22})
consists of two numbers $\lambda_{1}$ and $\lambda_{2}$ 
in the two down-spin sector.
Suppose that we consider only eq. (\ref{eq:BAE11}) and do not consider eq. (\ref{eq:BAE22}) for them.   
We replace $\lambda_{1}$ and $\lambda_{2}$ in eq. (\ref{eq:BAE11}) with rapidities $\mu_1$ and $\mu_2$, respectively, and assume that $\mu_1$ and $\mu_2$ satisfy eq. (\ref{eq:BAE11}) as continuous variables. Thus, we can regard rapidity $\mu_{2}$ as a function of rapidity $\mu_{1}$. 
Then, we shall assign another equation (\ref{eq:BAE22})   
on them in order to derive the solution of BAEs (\ref{eq:BAE11}) and (\ref{eq:BAE22}).

The first Bethe ansatz equation (\ref{eq:BAE11}) 
is equivalent to the following equation 


\begin{eqnarray}
\frac{\tan(\mu_{1}-\mu_{2})}{\tanh(\zeta)}=\frac{-1}{\tan\Bigl(N\tan^{-1}\bigl(\frac{\tan(\mu_{1})}{\tanh(\zeta/2)}\bigr)\Bigr)},\label{eq:lam2_express_lam1_3}
\end{eqnarray}

with the branch of the arc tangent function specified by the 
following inequalities  

\begin{eqnarray}
-\frac{\pi}{2}<N\tan^{-1}\biggl(\frac{\tan(\mu_{1})}{\tanh(\zeta/2)}\biggr)-\pi J_{1}-\pi\Bigl[\frac{2(\mu_{1}-\mu_{2})+\pi}{2\pi}\Bigr]_{Gauss}<\frac{\pi}{2}\nonumber . \\
\label{eq:lam2_express_lam1_1}
\end{eqnarray}

By making use of the addition theorem of the tangent function, equation (\ref{eq:lam2_express_lam1_3}) with  inequality (\ref{eq:lam2_express_lam1_1}) is equivalent to 
\begin{eqnarray}
\mu_{2}=\tan^{-1}\Biggl(-\frac{\frac{\tanh(\zeta)}{\tan\bigl(N\tan^{-1}\bigl(\frac{\tan(\mu_{1})}{\tanh(\zeta/2)}\bigr)\bigr)}+\tan(\mu_{1})}{\tan(\mu_{1})\frac{\tanh(\zeta)}{\tan\bigl(N\tan^{-1}\bigl(\frac{\tan(\mu_{1})}{\tanh(\zeta/2)}\bigr)\bigr)}-1}\Biggr)\label{eq:lambda2_with_lambda1}
\end{eqnarray}

where the range of rapidity $\mu_1$ is given by 

\begin{eqnarray}
-\frac{\pi}{2}&+\pi J_{1}+\pi\Bigl[\frac{2(\mu_{1}-\mu_{2})+\pi}{2\pi}\Bigr]_{Gauss}\nonumber\\
&<N\tan^{-1}\biggl(\frac{\tan(\mu_{1})}{\tanh(\zeta/2)}\biggr)<\frac{\pi}{2}+\pi J_{1}+\pi\Bigl[\frac{2(\mu_{1}-\mu_{2})+\pi}{2\pi}\Bigr]_{Gauss},
\end{eqnarray}

and we assume the following 


\begin{eqnarray}
\tan(\mu_{1})\frac{\tanh(\zeta)}{\tan\bigl(N\tan^{-1}\bigl(\frac{\tan(\mu_{1})}{\tanh(\zeta/2)}\bigr)\bigr)}-1\neq0.\label{eq:condition2}
\end{eqnarray}

Let us now consider discontinuity points for height function 
$h(\zeta, \mu_1)$ 
which is defined by eq. (\ref{eq:def_search_func}).
From equation (\ref{eq:def_search_func}), the candidates of the discontinuity points are 
$\mu_{1}-\mu_{2}=\frac{\pi}{2}$ and $\mu_{2}=\frac{\pi}{2}$. 
Here we remark that we do not consider $-\pi/2$ or $\pi$ since they are 
outside of the regions of (\ref{eq:regions}).
However, the point satisfying $\mu_{1}-\mu_{2}=\frac{\pi}{2}$ is not discontinuous since the change in Bethe quantum number balances it, although it seems to be discontinuous due to the term of the Gaussian symbol in height function $h(\zeta, \mu_1)$.  

Specifically, the Bethe quantum number and the term of the Gaussian sign change simultaneously around at this point, and their two changes cancel out each other 
so that function $h(\zeta, \mu_1)$ is continuous at this point, although each of them is discontinuous.  

%
%
Thus, hereafter we focus on the monotonicity of height function 
$h(\zeta, \mu_1)$ and the singularity point at 
$\mu_{2}=\frac{\pi}{2}$.

\section{Monotonicity and singular points of difference $\mu_2-\mu_1$
}\label{sec:monotonicity_of_the_height_function}

We shall show the monotonicity of height function 
$h(\zeta, \mu_1)$ as a function  of rapidity $\mu_1$ 
in this section.
First, we prove the monotonicity of the difference: 
$\mu_{2}-\mu_{1}$ as a function of rapidity $\mu_1$ 
except for the points where 
$\mu_{2}$ is equal to $\frac{\pi}{2}$.
Furthermore, we shall derive both the limit from the right and the limit from the left at each discontinuity point of the height function.

\subsection{
Monotonicity of difference $\mu_{2}-\mu_{1}$ as a  function of rapidity $\mu_1$}

We define the difference function  $P(\mu_{1})$ by $P(\mu_{1}):=\mu_{2}-\mu_{1}$ as a function of rapidity $\mu_1$. 
It is expressed as

\begin{eqnarray}
P(\mu_{1})&=\tan^{-1}\Biggl(-\frac{\frac{\tanh(\zeta)}{\tan\bigl(N\tan^{-1}\bigl(\frac{\tan(\mu_{1})}{\tanh(\zeta/2)}\bigr)\bigr)}+\tan(\mu_{1})}{\tan(\mu_{1})\frac{\tanh(\zeta)}{\tan\bigl(N\tan^{-1}\bigl(\frac{\tan(\mu_{1})}{\tanh(\zeta/2)}\bigr)\bigr)}-1}\Biggr)-\mu_{1}\nonumber\\
&=\tan^{-1}\bigl(f(\mu_{1})\bigr)-\mu_{1}, 
\label{eq:P(lambda)}
\end{eqnarray}

where 
$f(\mu_{1})$ is defined as follows.
\begin{eqnarray}
f(\mu_{1}):=-\frac{\frac{\tanh(\zeta)}{\tan\bigl(N\tan^{-1}\bigl(\frac{\tan(\mu_{1})}{\tanh(\zeta/2)}\bigr)\bigr)}+\tan(\mu_{1})}{\tan(\mu_{1})\frac{\tanh(\zeta)}{\tan\bigl(N\tan^{-1}\bigl(\frac{\tan(\mu_{1})}{\tanh(\zeta/2)}\bigr)\bigr)}-1}.
\end{eqnarray}

The derivative of the function 
$P(\mu_{1})$ with respect to 
$\mu_{1}$ is 
\begin{equation}
\frac{d}{d\mu_{1}}P(\mu_{1})=\frac{1}{1+f^{2}(\mu_{1})}\frac{df(\mu_{1})}{d\mu_{1}}-1\label{eq:deriv_p_1}.
\end{equation}

$f(\mu_{1})$ is expressed as
\begin{eqnarray}
f(\mu_{1})=\frac{a-b}{1+ab}
\end{eqnarray}
where $a$ and $b$ are given by
\begin{eqnarray}
a&:=\tan(\mu_{1})\\
b&:=-\frac{\tanh(\zeta)}{\tan\bigl(N\tan^{-1}\bigl(\frac{\tan(\mu_{1})}{\tanh(\zeta/2)}\bigr)\bigr)}.
\end{eqnarray}
We calculate the derivative of function 
$f(\mu_{1})$ as
\begin{eqnarray}
\frac{d}{d\mu_{1}}f(\mu_{1})&=\frac{(a'-b')(1+ab)-(a-b)(a'b+ab')}{(1+ab)^{2}}\nonumber\\
&=\frac{a'-b'-a^{2}b'+a'b^{2}}{(1+ab)^{2}}\label{eq:deri_f}
\end{eqnarray}
where the symbols $a'$ and $b'$ are given by 
\begin{eqnarray}
a'=\frac{d}{d\mu_{1}}a\\
b'=\frac{d}{d\mu_{1}}b.
\end{eqnarray}
From equation (\ref{eq:deri_f}), we show the derivative of function 
$P(\mu_{1})$ with respect to 
$\mu_{1}$ as follows.
\begin{eqnarray}
\frac{d}{d\mu_{1}}P(\mu_{1})&=\frac{1}{1+(\frac{a-b}{1+ab})^{2}}\frac{a'-b'-a^{2}b'+a'b^{2}}{(1+ab)^{2}}-1\nonumber\\
&=\frac{a'-b'-a^{2}b'+a'b^{2}-(1+ab)^{2}-(a-b)^{2}}{(1+ab)^{2}+(a-b)^{2}}\nonumber\\
&=\frac{(a'-a^{2})+(a'-a^{2})b^{2}-b'-a^{2}b'-1-b^{2}}{(1+ab)^{2}+(a-b)^{2}}.\label{eq:deriv_p_2}
\end{eqnarray}
Here, we remark
\begin{eqnarray}
a'-a^{2}&=\frac{1}{\cos^{2}(\mu_{1})}-\frac{\sin^{2}(\mu_{1})}{\cos^{2}(\mu_{1})}\nonumber\\
&=1.
\end{eqnarray}
Here we remark that the derivative of $b$ is positive:    $b'>0$. Thus, it follows that the derivative of the difference function $P$ is negative as follows.
\begin{eqnarray}
\frac{d}{d\mu_{1}}P(\mu_{1})&=-\frac{1+a^{2}}{(1+ab)^{2}+(a-b)^{2}}b'\nonumber\\
&<0.
\end{eqnarray}
Thus, we have shown that function 
$P(\mu_{1})=\mu_{2}-\mu_{1}$ is monotonically decreasing.

\subsection{Limiting values of difference $\mu_2-\mu_1$ at the singular points of the height function}

We define constant number $K^{J_{1}}$ as the solution $\mu_{1}$ of the equation
\begin{eqnarray}
\tan(\mu_{1})\frac{\tanh(\zeta)}{\tan\bigl(N\tan^{-1}\bigl(\frac{\tan(\mu_{1})}{\tanh(\zeta/2)}\bigr)\bigr)}-1=0
\label{eq:def_KJ}
\end{eqnarray}
such that it gives the infimum of the interval  
\begin{equation}
\tan^{-1}\bigl(\tanh(\zeta/2)\tan\bigl(\frac{\pi}{N}(J_{1} 
\textcolor{red}{-} \frac{1}{2})\bigr)\bigr) < \mu_{1} < \tan^{-1}\bigl(\tanh(\zeta/2)\tan\bigl(\frac{\pi}{N}(J_{1}+ 
\textcolor{red}{\frac{1}{2}} )\bigr)\bigr). 
\label{eq:range}
\end{equation}

\begin{figure}[ht]
  \centering
  \includegraphics[keepaspectratio, scale=0.5]{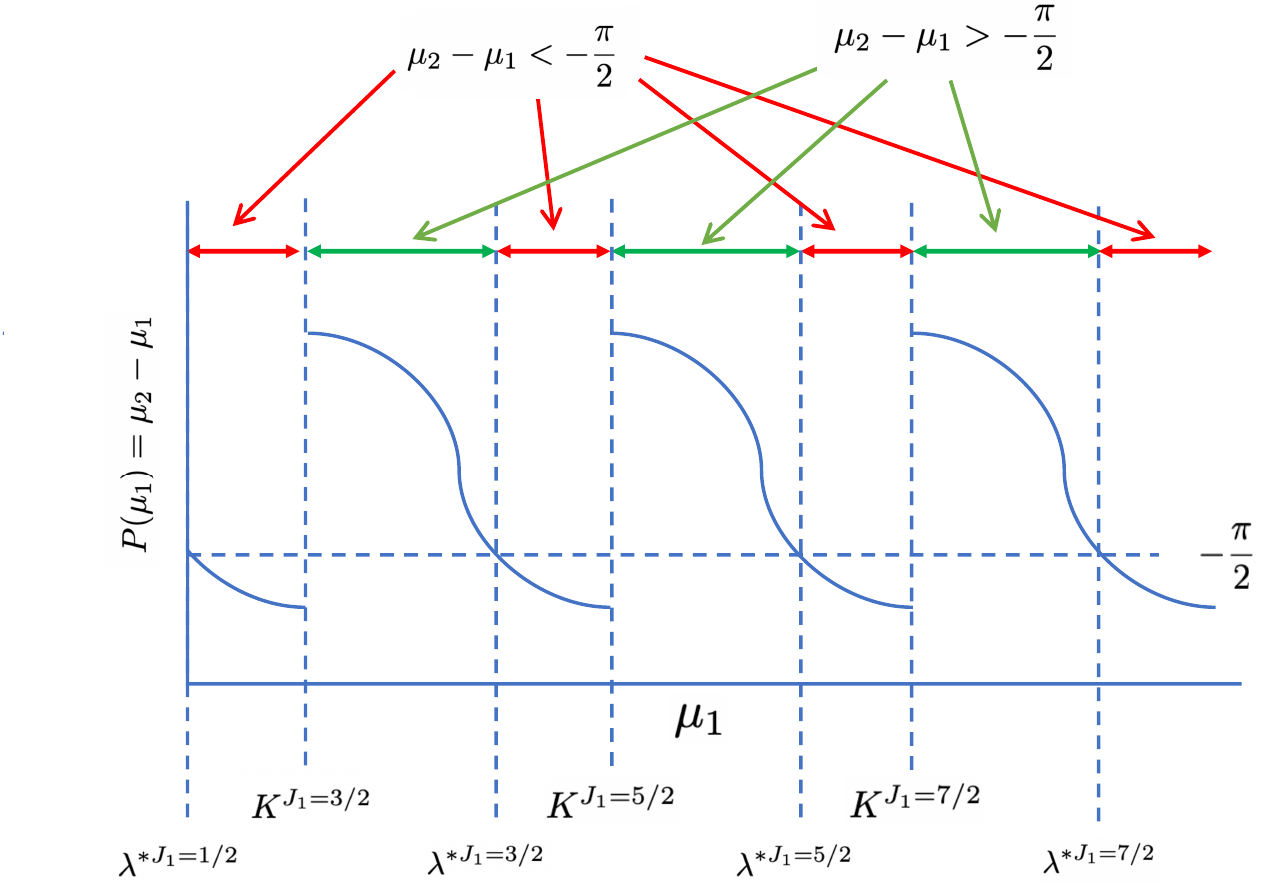}
  \caption{ 
Illustration of the graph
of $P(\mu_1)$ versus $\mu_1$.
Function $P(\mu_1)$ is discontinuous at $\mu_1=K^{J_{1}}$ with respect to $\mu_1$. Due to the constraint (\ref{eq:lam2_express_lam1_1}), $J_1$, one of the Bethe quantum numbers corresponding to the Bethe solutions $\lambda_{1}$ and $\lambda_{2}$, changes to $J_1+1$ at $\mu_1=K^{J_1+1}$ as $\mu_1$ increases.  Function $P(\mu_1)$  is monotonically decreasing in the interval from $K^{J_1}$ to $K^{J_1+1}$ as analytically shown in section 4.1.  
$P(\mu_1)$ equals to $-\pi/2$ at $\mu_1=\lambda_1^{*J_1}$ for each value of $J_1$:  
$P(\mu_1=\lambda_1^{*J_1})=-\pi/2$.
The domains of rapidity $\mu_1$ in eqs. (\ref{eq:domain1}) and 
(\ref{eq:domain2}) are expressed with green and red lines, 
respectively, and they correspond to the parts of the graph $P(\mu_1)<-\pi/2$ and $P(\mu_1)>-\pi/2$, 
respectively. Green lines correspond to the domains (\ref{eq:domain1}) of $\mu_1 - \mu_2 < \pi/2$.
} 
\label{lambda_2-lambda_1_expand}
\end{figure}

Here we remark that the condition of $\mu_1 - \mu_2 > \pi/2$ in eq. (\ref{eq:domain2}) 
corresponds to $\mu_2 - \mu_1 < - \pi/2$
in Figure \ref{lambda_2-lambda_1_expand}.

We next define constant number $\lambda^{*J_{1}}_{1}$ by
\begin{eqnarray}
\lambda_{1}^{*J_{1}}:=\tan^{-1}\bigl(\tanh(\zeta/2)\tan\bigl(\frac{\pi}{N}(J_{1} + \frac{1}{2})\bigr)\bigr)
\end{eqnarray}
By taking the limit of $\mu_{1}$ approaching $\lambda_{1}^{*J_{1}}$, we can show  
\begin{eqnarray}
\lim_{\mu_{1} \rightarrow \lambda_{1}^{*J_{1}}} P(\mu_{1})=-\frac{\pi}{2}
 \label{eq:p_limit_j1+1/2},
\end{eqnarray}
We derive it as follows. 
First, we show that $\tan(N \tan^{-1}(\tan(\mu_1))/\tanh(\zeta/2))$ approaches zero in the limit of sending $\mu_1$ to $\lambda_1^{*J_1}$. 
Then, we substitute the infinitesimally small value approaching zero in eq. 
(\ref{eq:P(lambda)}), and we obtain the difference $- \pi/2$ 
by making use of the following formula:  
$1/\tan \theta = \cot \theta = \pi/2 - \tan \theta$.

Here we remark that $K^{J_1}$ corresponds to an upper bound to the two domains associated with $J_1$, while $\lambda_1^{*J_1}$ is located between the two domains associated with $J_1$, as shown in Fig. \ref{lambda_2-lambda_1_expand}.

We now calculate function $\tan(\mu_2(\mu_1))$ 
through the limit of sending $\mu_{1}$ to $K^{J_{1}}$ with $\mu_{1} < K^{J_{1}}$, i.e., from the left. 
We also calculate it through the limit of sending $\mu_{1}$ to $K^{J_{1}}$ with $\mu_{1} > K^{J_{1}}$, i.e., from the right, which is denoted by $\mu_{1} \downarrow K^{J_{1}}$, as follows.   
\begin{eqnarray}
\lim_{\mu_{1}\uparrow K^{J_{1}}}\tan(\mu_{2}(\mu_{1}))=-\infty\label{eq:limit_K0_uparrow},\\
\lim_{\mu_{1}\downarrow K^{J_{1}}}\tan(\mu_{2}(\mu_{1}))=\infty.\label{eq:limit_K0_downarrow}
\end{eqnarray}
It follows that 
\begin{eqnarray}
    \lim_{\mu_{1}\uparrow K^{J_{1}}}\mu_{2}(\mu_{1})=- \pi/2,
    \label{eq:limit_K0_uparrow2}
    \\
\lim_{\mu_{1}\downarrow K^{J_{1}}} \mu_{2}(\mu_{1})=\pi/2.
\label{eq:limit_K0_downarrow2}
\end{eqnarray}
We therefore obtain the following relations.
\begin{eqnarray} \lim_{\mu_{1}\uparrow K^{J_{1}}}P(\mu_{1})<-\frac{\pi}{2},\label{eq:p_limit_lambda1}\\
\lim_{\mu_{1}\downarrow K^{J_{1}}}P(\mu_{1})>-\frac{\pi}{2}.\label{eq:p_limit_lambda2} \end{eqnarray}
Here we have derived eq. (\ref{eq:p_limit_lambda1}) since $P(\mu_1) = \mu_2 - \mu_1$ and $\mu_1 >0$, 
and  eq. (\ref{eq:p_limit_lambda2}) since $P(\mu_1) = \pi/2 - \mu_1$ and $\mu_1 < \pi/2$.

\subsection{Graphical illustration of difference   
$P(\mu_{1})=\mu_{2}(\mu_1)-\mu_{1}$ 
as a function of rapidity $\mu_1$
}
%
We recall that 
$P(\mu_{1})=\mu_{2}-\mu_{1}$ is monotonically decreasing except for the point 
$\mu_{1}=K^{J_{1}}$.
We also recall inequalities (\ref{eq:p_limit_lambda1}) and (\ref{eq:p_limit_lambda2}).
It follows that
in the range 
$\lambda_{1}^{*J_{1}}<\mu_{1}<K^{J_{1}}$, 
we obtain
\begin{eqnarray}
P(\mu_{1})>-\frac{\pi}{2}.
\end{eqnarray}
Similarly for the range 
$K^{J_{1}}<\mu_{1}<\lambda_{1}^{*J_{1}+1}$, we obtain 
\begin{eqnarray}
P(\mu_{1})<-\frac{\pi}{2}.
\end{eqnarray}
From the monotonicity and the limit from the right and the limit from the left of 
$\mu_{2}-\mu_{1}$ at the point satisfying 
$\mu_{1} = K^{J_1}$ (see (\ref{eq:p_limit_lambda1}) and (\ref{eq:p_limit_lambda2})), we obtain the shape of 
$P(\mu_{1})=\mu_{2}-\mu_{1}$ in figure \ref{lambda_2-lambda_1_expand}.
From this figure, it follows that
$P(\mu_{1})$ is equal to $-\pi/2$ at 
$\mu_{1}=\lambda_{1}^{*J_{1}}$.
In addition, we obtain the relation that 
$P(\mu_1)$ is discontinuous at 
$\mu_1=K^{J_1}$.
It is also monotonically decreasing in the interval from $K^{J_1}$ to $K^{J_1+1}$.
From these properties, we derive following relations; 
$P(\mu_1)<-\pi/2$ in 
$\mu_{1}\geq\lambda_{1}^{*J_{1}}$ and $P(\mu_1)>-\pi/2$ in 
$\mu_{1}<\lambda_{1}^{*J_{1}}$.

\section{The analysis of discontinuity in height function 
$h(\zeta, \mu_{1})$}\label{sec:h_func}

We recall that the second Bethe ansatz equation (\ref{eq:BAE22}) is expressed in terms of the height function (\ref{eq:def_search_func}) as 
\begin{eqnarray}
J_{2}=h(\zeta, \mu_{1})\label{eq:BAE_height_function}.
\end{eqnarray}
When two quantum numbers are different: $J_{1}\neq J_{2}$, 
we derive the Bethe quantum numbers by using 
$h(\zeta, \mu_{1})$, in this section.
From the monotonicity of height function 
$h(\zeta, \mu_{1})$ except for the points where 
$\mu_{1}$ is equal to $K^{J_{1}}$, in order to obtain the Bethe quantum numbers, we derive the limit from the right and the limit from the left of height function 
$h(\zeta, \mu_{1})$.
In addition, we derive the limit of height function 
$h(\zeta, \mu_{1})$ at the 
$\mu_{1}=\frac{\pi}{2}$.

We remark that the case $J_{1}=J_{2}$ cannot be dealt with by this approach.
The way to solve the Bethe ansatz equation for $J_{1}=J_{2}$ using the counting function is described in section \ref{sec:intro_counting} and \ref{sec:collapse_extra}.

\subsection{Enumeration of real solutions when $J_{1}\neq J_{2}$}

From the analysis of height function 
$h(\zeta, \mu_{1})$ for each $J_{1}$, we obtain all the Bethe quantum numbers for real solutions.
The Bethe quantum numbers $J_{1}$ and $J_{2}$ except for the $J_{2}=J_{1}$ case satisfy
\begin{eqnarray}
-\frac{N-1}{2}<J_{1}<J_{2}<\frac{N-1}{2}\label{eq:finite_bqn_set}
\end{eqnarray}
and for $J_{2}=(N-1)/2$
\begin{eqnarray}
(J_{1}, J_{2}) = \Bigl(\frac{1}{2}, \frac{N-1}{2}\Bigr), \Bigl(\frac{3}{2}, \frac{N-1}{2}\Bigr),\cdots , \Bigl(\frac{N-3}{2}, \frac{N-1}{2}\Bigr).\label{eq:J1J2_equation}
\end{eqnarray}

We derive the list of the Bethe quantum number for real solutions (\ref{eq:finite_bqn_set}) and (\ref{eq:J1J2_equation}) in the following subsections.

\subsection{Limits of height function 
$h(\zeta, \mu_{1})$}

In the limit as 
$\mu_{1}$ approaches 
$K^{J_{1}}$ for $\mu_{1}$ with $K^{J_{1}}<\mu_{1}$, we evaluate function 
$h(\zeta, \mu_{1})$.
\begin{eqnarray}
\lim_{\mu_{1}\downarrow K^{J_{1}}}h(\zeta,\mu_{1})&=\frac{N}{\pi}\cdot\frac{\pi}{2}-\frac{1}{\pi}\tan^{-1}\biggl(\frac{1}{\tanh(\zeta)}\Bigl(\frac{1}{\tan(K^{J_{1}})}\Bigr)\biggr)\nonumber\\
&>\frac{N}{2}-\frac{1}{\pi}\cdot\frac{\pi}{2}\nonumber\\
&=\frac{N-1}{2}
\end{eqnarray}

Similarly, in the limit as 
$\mu_{1}$ approaches $K^{J_{1}}$ for 
$\mu_{1}$ with 
$\mu_{1}<K^{J_{1}}$, we evaluate function 
$h(\zeta, \mu_{1})$.
\begin{eqnarray}
\lim_{\mu_{1}\uparrow K^{J_1}}h(\zeta,\mu_{1})&=-\frac{N}{\pi}\cdot\frac{\pi}{2}-\frac{1}{\pi}\tan^{-1}\biggl(\frac{1}{\tanh(\zeta)}\Bigl(\frac{1}{\tan(K^{J_{1}})}\Bigr)\biggr)+1\nonumber\\
&<-\frac{N}{2}+\frac{1}{\pi}\frac{\pi}{2}+1\nonumber\\
&=-\frac{N-3}{2}
\end{eqnarray}
Thus, we derive the graph of function 
$h(\zeta, \mu_{1})$ in Figure \ref{fig:h(zeta)_one}.

\subsection{The value of 
$\dsp\lim_{\mu_{1}\rightarrow\frac{\pi}{2}}h(\zeta,\mu_{1})$ when $J_{1}=\dsp\frac{N-1}{2}$ and $N$ is even}\label{eq:h_limit_pi2_even}
In this subsection, when the site number $N$ is even, we calculate 
$h(\zeta, \mu_{1})$ by sending 
$\mu_{1}$ to $\frac{\pi}{2}$. 
First, we remark
\begin{eqnarray}
\tan{(\mu_{2})}(\mu_{1})&=&\frac{\tan{(\mu_{1})}+\tanh(\zeta)\tan(N\tan^{-1}(\frac{\tan(\mu_{1})}{\tanh\zeta/2})-\pi J_{1})}{1+\tanh(\zeta)\tan{(\mu_{1})}\tan(N\tan^{-1}(\frac{\tan(\mu_{1})}{\tanh\zeta/2})-\pi J_{1})}\nonumber\\
&=&\frac{\tan{(\mu_{1})}+\tanh(\zeta)\tan(N\tan^{-1}(\frac{\tan(\mu_{1})}{\tanh\zeta/2})-\pi/2)}{1+\tanh(\zeta)\tan{(\mu_{1})}\tan(N\tan^{-1}(\frac{\tan(\mu_{1})}{\tanh\zeta/2})-\pi/2)}.\nonumber\\
\end{eqnarray}
Here since $N$ is even, we can show
\begin{eqnarray}
\tan\biggl(N\tan^{-1}\Bigl(\frac{\tan(\mu_{1})}{\tanh(\zeta/2)}\Bigr)-\pi/2\biggr)\rightarrow\infty \ (\tan(\mu_{1})\rightarrow\infty).\label{eq:tan_lambda2_limit_pi2_even}
\end{eqnarray}
From the equation (\ref{eq:tan_lambda2_limit_pi2_even}), we obtain
\begin{eqnarray}
\tan(\mu_{2})(\mu_{1})=\frac{\frac{1}{\tan(N\tan^{-1}(\frac{\tan(\mu_{1})}{\tanh\zeta/2})-\pi/2)}-\frac{\tanh(\zeta)}{\tan(\mu_{1})}}{\frac{1}{\tan(\mu_{1})\tan(N\tan^{-1}(\frac{\tan(\mu_{1})}{\tanh\zeta/2})-\pi/2)}+\tanh(\zeta)}\nonumber\\
\rightarrow0\ \ \ (\tan(\mu_{1})\rightarrow\infty).
\end{eqnarray}
Thus, we have 
\begin{eqnarray}
\lim_{\mu_{1}\rightarrow\frac{\pi}{2}}h(\zeta,\mu_{1})=\frac{1}{2}.\label{eq:height_fuc_pi/2lim_even}
\end{eqnarray}

\subsection{The value of 
$\dsp \lim_{\mu_{1}\rightarrow\frac{\pi}{2}}h(\zeta,\mu_{1})$ when $J_{1}=\dsp\frac{N-1}{2}$ and $N$ is odd}\label{eq:h_limit_pi2_odd}
In this subsection, when the site number $N$ is odd, we calculate 
$h(\zeta, \mu_{1})$ by sending 
$\mu_{1}$ to $\frac{\pi}{2}$. 
We remark
\begin{eqnarray}
\tan{(\mu_{2})}(\mu_{1})&=&\frac{\tan{(\mu_{1})}+\tanh(\zeta)\tan(N\tan^{-1}(\frac{\tan(\mu_{1})}{\tanh\zeta/2})-\pi J_{1})}{1+\tanh(\zeta)\tan{(\mu_{1})}\tan(N\tan^{-1}(\frac{\tan(\mu_{1})}{\tanh\zeta/2})-\pi J_{1})}\nonumber\\
&=&\frac{\tan{(\mu_{1})}+\tanh(\zeta)\tan(N\tan^{-1}(\frac{\tan(\mu_{1})}{\tanh\zeta/2}))}{1+\tanh(\zeta)\tan{(\mu_{1})}\tan(N\tan^{-1}(\frac{\tan(\mu_{1})}{\tanh\zeta/2}))}.
\end{eqnarray}
Here from $N$ is odd, we remark 
\begin{eqnarray}
\tan\biggl(N\tan^{-1}\Bigl(\frac{\tan(\mu_{1})}{\tanh(\zeta/2)}\Bigr)\biggr)\rightarrow\infty \ (\tan(\mu_{1})\rightarrow\infty).\label{eq:tan_lambda2_limit_pi2_odd}
\end{eqnarray}
From the equation (\ref{eq:tan_lambda2_limit_pi2_odd}), we derive
\begin{eqnarray}
\tan(\mu_{2})(\mu_{1})&=&\frac{\frac{1}{\tan(N\tan^{-1}(\frac{\tan(\mu_{1})}{\tanh\zeta/2}))}-\frac{\tanh(\zeta)}{\tan(\mu_{1})}}{\frac{1}{\tan(\mu_{1})\tan(N\tan^{-1}(\frac{\tan(\mu_{1})}{\tanh\zeta/2}))}+\tanh(\zeta)}\nonumber\\
&\rightarrow&0\ \ \ (\tan(\mu_{1})\rightarrow\infty).
\end{eqnarray}
Thus, we obtain
\begin{eqnarray}
\lim_{\mu_{1}\rightarrow\frac{\pi}{2}}h(\zeta,\mu_{1})=\frac{1}{2}.\label{eq:height_fuc_pi/2lim_odd}
\end{eqnarray}

\subsection{In the case when $J_{1}=J_{2}$}

When $J_{1}=J_{2}$, it is not straightforward to derive a real solution of the Bethe ansatz equations by making use of the height function.
When the two quantum numbers are equal: $J_{1}=J_{2}$, there are two types of solutions to Bethe ansatz equations (\ref{eq:BAE11}) and (\ref{eq:BAE22}). 
The first case gives a trivial solution with $\lambda_{1}=\lambda_{2}$, which leads to the following Bethe ansatz equation:
\begin{eqnarray}
2\tan^{-1}\biggl(\frac{\tan(\lambda_{1})}{\tanh(\frac{\zeta}{2})}\biggr)=\frac{2\pi}{N}J_{1}.
\end{eqnarray} 
In this case, however, the solution to the Bethe ansatz equations doesn't correspond to a valid quantum state. 

The second type of solution is 
$\lambda_{1} \neq \lambda_{2}$. 
There are two cases $(\lambda_{1}, \lambda_{2})=(\tilde{\lambda}_{1}, \tilde{\lambda}_{2}),(\tilde{\lambda}_{2}, \tilde{\lambda}_{1})$ where $\tilde{\lambda}_{1}$ and $\tilde{\lambda}_{2}$ are labeled with the values of the solutions of the Bethe ansatz equation in this case. 
These cases correspond to the same state.

Thus, in this case, if there exist three solutions of the Bethe ansatz equations, the Bethe quantum number $(J_{1}, J_{2})$ that satisfy $J_{1}=J_{2}$ has the real solution corresponding to the valid quantum state. 
The Bethe quantum numbers for the case $J_{1}=J_{2}$ are obtained in section 6 and section 7 using the counting function approach.

\section{The counting functions for real solutions in the two down spin sector}\label{sec:intro_counting}

In the previous section, we have derived all Bethe quantum numbers for
real solutions associated with two different Bethe quantum numbers 
(i.e., $J_1 \ne J_2$). 
In this section, we analyze the Bethe quantum numbers for real solutions with $J_1=J_2$.
In subsection \ref{sec:Deriv_count_function}, we introduce the counting function. In subsection \ref{sec:repre_tan^2}, We derive $\tan^{2}(x)$ where $x$ is the 
center of two rapidities of a real solution. In subsection \ref{sec:massive_real_quantum}, we derive 
all the integers which appear as the difference of the two Bethe quantum numbers.

\subsection{
Counting function for real solutions}\label{sec:Deriv_count_function}


We now recall the Bethe ansatz equations (\ref{eq:BAE11}) and (\ref{eq:BAE22}) 
in the two down-spin sector.
We express real solutions of the Bethe ansatz equation in the two down-spin sector as follows.
\begin{eqnarray}
\lambda_{1}&=x-\frac{1}{2}\gamma\zeta\label{eq:real_sol_1}\\
\lambda_{2}&=x+\frac{1}{2}\gamma\zeta\label{eq:real_sol_2}
\end{eqnarray}
Here we call $x$ the center and $\gamma \zeta$  the deviation of a real solution in the two down-spin sector. 
We now calculate each term in the first Bethe ansatz equation (\ref{eq:BAE11}). By using the 
expressions (\ref{eq:real_sol_1}) and (\ref{eq:real_sol_2}), we calculate the left-hand side of the first Bethe ansatz equation (\ref{eq:BAE11}) as follows. 
\begin{eqnarray}
2&\tan^{-1}\biggl(\frac{\tan(\lambda_{1})}{\tanh(\frac{\zeta}{2})}\biggr)=2\tan^{-1}\biggl(\frac{\tan(x-\frac{1}{2}\gamma\zeta)}{\tanh(\frac{\zeta}{2})}\biggr)\nonumber\\
&=\frac{1}{i}\biggl\{\log\biggl(1+i\frac{\tan(x-\frac{1}{2}\gamma\zeta)}{\tanh(\frac{\zeta}{2})}\biggr)-\log\biggl(1-i\frac{\tan(x-\frac{1}{2}\gamma\zeta)}{\tanh(\frac{\zeta}{2})}\biggr)\biggr\}\nonumber\\
&=\frac{1}{i}\log\biggl\{\frac{\tanh(\frac{\zeta}{2})+i\tan(x-\frac{1}{2}\gamma\zeta)}{\tanh(\frac{\zeta}{2})-i\tan(x-\frac{1}{2}\gamma\zeta)}\biggr\}\nonumber\\
&=\frac{1}{2i}\log\biggl\{\frac{\tanh(\frac{\zeta}{2})+i\tan(x-\frac{1}{2}\gamma\zeta)}{\tanh(\frac{\zeta}{2})-i\tan(x-\frac{1}{2}\gamma\zeta)}\biggr\}+\frac{1}{2i}\log\biggl\{\frac{\tanh(\frac{\zeta}{2})+i\tan(x+\frac{1}{2}\gamma\zeta)}{\tanh(\frac{\zeta}{2})-i\tan(x+\frac{1}{2}\gamma\zeta)}\biggr\}\nonumber\\
&+\frac{1}{2i}\log\biggl\{\frac{\tanh(\frac{\zeta}{2})+i\tan(x-\frac{1}{2}\gamma\zeta)}{\tanh(\frac{\zeta}{2})-i\tan(x-\frac{1}{2}\gamma\zeta)}\biggr\}-\frac{1}{2i}\log\biggl\{\frac{\tanh(\frac{\zeta}{2})+i\tan(x+\frac{1}{2}\gamma\zeta)}{\tanh(\frac{\zeta}{2})-i\tan(x+\frac{1}{2}\gamma\zeta)}\biggr\}\nonumber\\
&=\tan^{-1}\biggl(\frac{\tan(x-\frac{1}{2}\gamma\zeta)}{\tanh(\frac{\zeta}{2})}\biggr)+\tan^{-1}\biggl(\frac{\tan(x+\frac{1}{2}\gamma\zeta)}{\tanh(\frac{\zeta}{2})}\biggr)\nonumber\\
&+\frac{1}{2i}\log\biggl\{\frac{(\tanh(\frac{\zeta}{2})+i\tan(x-\frac{1}{2}\gamma\zeta))(\tanh(\frac{\zeta}{2})-i\tan(x+\frac{1}{2}\gamma\zeta))}{(\tanh(\frac{\zeta}{2})-i\tan(x-\frac{1}{2}\gamma\zeta))(\tanh(\frac{\zeta}{2})+i\tan(x+\frac{1}{2}\gamma\zeta))}\biggr\}.
\end{eqnarray}
By using expressions (\ref{eq:real_sol_1}) and (\ref{eq:real_sol_2}), we calculate the second term of the right-hand side of the first Bethe ansatz equation (\ref{eq:BAE11}) as follows. 
\begin{eqnarray}
\frac{2}{N}\tan^{-1}&\biggl(\frac{\tan(\lambda_{1}-\lambda_{2})}{\tanh{\zeta}}\biggr)=\frac{2}{N}\tan^{-1}\biggl(\frac{\tan(-\gamma\zeta)}{\tanh(\zeta)}\biggr)\nonumber\\
&=\frac{1}{N}\biggl\{\frac{1}{i}\log\biggl(1+i\frac{\tan(-\gamma\zeta)}{\tanh(\zeta)}\biggr)-\frac{1}{i}\log\biggl(1-i\frac{\tan(-\gamma\zeta)}{\tanh(\zeta)}\biggr)\biggr\}\nonumber\\
&=\frac{1}{N}\biggl\{\frac{1}{i}\log\biggl(\frac{\tanh(\zeta)+i\tan(-\gamma\zeta)}{\tanh(\zeta)-i\tan(-\gamma\zeta)}\biggr)\biggr\}\nonumber\\
&=\frac{1}{N}\biggl\{\frac{1}{2i}\log\biggl(\frac{\bigl(\tanh(\zeta)+i\tan(-\gamma\zeta)\bigr)^2}{\bigl(\tanh(\zeta)-i\tan(-\gamma\zeta)\bigr)^2}\biggr)\biggr\}.\label{eq:real_part_left_first_term}
\end{eqnarray}

We now introduce a function of two variables, i.e.,  center $x$ and deviation $\gamma$, denoted by $W(x, \gamma, \zeta)$. Here we recall that they express rapidity $\mu_1$ in eq. (\ref{eq:real_sol_1}).  
We define it by
\begin{eqnarray}
2\pi W(x,\gamma,\zeta)\equiv \tan^{-1}\biggl(\frac{\tan(x-\frac{1}{2}\gamma\zeta)}{\tanh(\frac{\zeta}{2})}\biggr)+\tan^{-1}\biggl(\frac{\tan(x+\frac{1}{2}\gamma\zeta)}{\tanh(\frac{\zeta}{2})}\biggr)-\frac{2\pi}{N}\biggl[\frac{-2\gamma\zeta+\pi}{2\pi}\biggr]_{\mathrm{Gauss}}.\nonumber\\
\label{eq:real_part_right_second_term}
\end{eqnarray}
We call $W(x, \gamma, \zeta)$ the counting function. 
Here we remark that 
the counting function (\ref{eq:real_part_right_second_term}) 
is expressed with the height function $h(\zeta, \lambda_{1})$ from (\ref{eq:def_search_func}) 
as 
\begin{eqnarray}
2\pi W(x,\gamma,\zeta)=\frac{\pi}{N}h\Bigl(\zeta, x- {\frac 1 2} \gamma \zeta \Bigr)\label{eq:BAE_counting_function_real}.
\end{eqnarray}
Thus, they are equivalent. However, the different choice of parameters such as the center $x$ with deviation $\gamma \zeta$ than rapidities $\mu_1$ and $\mu_2$ plays a central role when we solve the Bethe equations for the case of equal Bethe quantum numbers: $J_1=J_2$.
 
It follows from the above calculations (\ref{eq:real_part_left_first_term}) and (\ref{eq:real_part_right_second_term}), the first Bethe ansatz equation (\ref{eq:BAE11}) is expressed in terms of the counting function (\ref{eq:real_part_right_second_term}) as follows
\begin{eqnarray}
\frac{2\pi}{N}J_{1}&\equiv&2\tan^{-1}\biggl(\frac{\tan(\lambda_{1})}{\tanh(\frac{\zeta}{2})}\biggr)-\frac{2}{N}\tan^{-1}\biggl(\frac{\tan(\lambda_{1}-\lambda_{2})}{\tanh{\zeta}}\biggr)-\frac{2\pi}{N}\biggl[\frac{-2\gamma\zeta+\pi}{2\pi}\biggr]_{\mathrm{Gauss}}\nonumber\\
&=&2\pi W(x,\gamma,\zeta)+\frac{1}{2i}\log\biggl\{\frac{\bigl(\tanh(\frac{\zeta}{2})+i\tan(x-\frac{1}{2}\gamma\zeta)\bigr)\bigl(\tanh(\frac{\zeta}{2})-i\tan(x+\frac{1}{2}\gamma\zeta)\bigr)}{\bigl(\tanh(\frac{\zeta}{2})-i\tan(x-\frac{1}{2}\gamma\zeta)\bigr)\bigl(\tanh(\frac{\zeta}{2})+i\tan(x+\frac{1}{2}\gamma\zeta)\bigr)}\biggr\}\nonumber\\
&&-\frac{1}{N}\biggl\{\frac{1}{2i}\log\biggl(\frac{\bigl(\tanh(\zeta)+i\tan(-\gamma\zeta)\bigr)^2}{\bigl(\tanh(\zeta)-i\tan(-\gamma\zeta)\bigr)^2}\biggr)\biggr\}\label{eq:CF1}
\end{eqnarray}
Next, we calculate the imaginary part of the equation (\ref{eq:CF1}). 
At first, we confirm the next relation as follows.
We first note the following relation holds for an integer $n$:  
\begin{eqnarray}
\frac{1}{N}\biggl\{\log\biggl(\frac{\bigl(\tanh(\zeta)+i\tan(-\gamma\zeta)\bigr)^2}{\bigl(\tanh(\zeta)-i\tan(-\gamma\zeta)\bigr)^2}\biggr)\biggr\}=\log\biggl\{\exp\biggl(\frac{2\pi in}{N}\biggr)\biggl(\frac{\bigl(\tanh(\zeta)+i\tan(-\gamma\zeta)\bigr)^2}{\bigl(\tanh(\zeta)-i\tan(-\gamma\zeta)\bigr)^2}\biggr)^{1/N}\biggr\}\nonumber\\
\end{eqnarray}
The imaginary part of eq. (\ref{eq:CF1}) vanishes if the following relation holds for an integer $n$ with $n=0,1,\cdots,N-1$: 
\begin{eqnarray}
\Biggl\{\frac{\bigl(\tanh(\frac{\zeta}{2})+i\tan(x-\frac{1}{2}\gamma\zeta)\bigr)\bigl(\tanh(\frac{\zeta}{2})-i\tan(x+\frac{1}{2}\gamma\zeta)\bigr)}{\bigl(\tanh(\frac{\zeta}{2})-i\tan(x-\frac{1}{2}\gamma\zeta)\bigr)\bigl(\tanh(\frac{\zeta}{2})+i\tan(x+\frac{1}{2}\gamma\zeta)\bigr)}\Biggr\}\nonumber\\
=\exp\biggl(\frac{2\pi in}{N}\biggr)\biggl(\frac{\bigl(\tanh(\zeta)+i\tan(-\gamma\zeta)\bigr)^2}{\bigl(\tanh(\zeta)-i\tan(-\gamma\zeta)\bigr)^2}\biggr)^{\frac{1}{N}}\label{eq:Im}
\end{eqnarray}
It thus follows that the first Bethe ansatz equation (\ref{eq:BAE11}) holds 
if eq. (\ref{eq:Im}) is valid and the following equation holds
for an integer $n$ with $n=0, -1,\cdots,-(N-1)$: 
\begin{eqnarray}
\frac{2\pi}{N}J_{1}&=&2\pi W(x,\gamma,\zeta)+\frac{1}{2i}\log\exp\biggl(\frac{2\pi in}{N}\biggr)\nonumber\\
&=&2\pi W(x,\gamma,\zeta)+\frac{1}{2i}\biggl(\frac{2\pi in}{N}\biggr)\nonumber\\
&=&2\pi W(x,\gamma,\zeta)+\frac{\pi n}{N}\label{eq:real_counting_function_XXZ}
\end{eqnarray}
Making use of eq. (\ref{eq:Im}), we shall express $\tan{x}$ as a function of $\gamma$ as follows. 
\begin{eqnarray}
\biggl(&&\mbox{LHS of (\ref{eq:Im})}\biggr)\nonumber\\
&&=\frac{\tanh^2(\frac{\zeta}{2})-i\tanh(\frac{\zeta}{2})\Bigl(\tan(x+\frac{1}{2}\gamma\zeta)-\tan(x-\frac{1}{2}\gamma\zeta)\Bigr)+\tan(x-\frac{1}{2}\gamma\zeta)\tan(x+\frac{1}{2}\gamma\zeta)}{\tanh^2(\frac{\zeta}{2})+i\tanh(\frac{\zeta}{2})\Bigl(\tan(x+\frac{1}{2}\gamma\zeta)-\tan(x-\frac{1}{2}\gamma\zeta)\Bigr)+\tan(x-\frac{1}{2}\gamma\zeta)\tan(x+\frac{1}{2}\gamma\zeta)}\label{tan_im}\nonumber\\
\end{eqnarray}
We remark the following two relations.
\begin{eqnarray}
\tan\Bigl(x+\frac{1}{2}\gamma\zeta\Bigr)-\tan\Bigl(x-\frac{1}{2}\gamma\zeta\Bigr)&=\frac{\tan(x)+\tan(\frac{1}{2}\gamma\zeta)}{1-\tan(x)\tan(\frac{1}{2}\gamma\zeta)}-\frac{\tan(x)-\tan(\frac{1}{2}\gamma\zeta)}{1+\tan(x)\tan(\frac{1}{2}\gamma\zeta)}\nonumber\\
&=\frac{2\tan^2(x)\tan(\frac{1}{2}\gamma\zeta)+2\tan(\frac{1}{2}\gamma\zeta)}{1-\tan^2(x)\tan^2(\frac{1}{2}\gamma\zeta)}\label{eq:remark_tanx_two_relation1}
\end{eqnarray}
and
\begin{eqnarray}
\tan\Bigl(x+\frac{1}{2}\gamma\zeta\Bigr)\tan\Bigl(x-\frac{1}{2}\gamma\zeta\Bigr)=\frac{\tan^2(x)-\tan^2(\frac{1}{2}\gamma\zeta)}{1-\tan^2(x)\tan^2(\frac{1}{2}\gamma\zeta)}\label{eq:remark_tanx_two_relation2}
\end{eqnarray}
It follows from eqs. (\ref{eq:remark_tanx_two_relation1}) and (\ref{eq:remark_tanx_two_relation2}) 
that eq. (\ref{tan_im}) is expressed as
\begin{eqnarray}
&\frac{\tanh^2\Bigl(\frac{\zeta}{2}\Bigr)-i\tanh\Bigl(\frac{\zeta}{2}\Bigr)\Bigl(\frac{2\tan^2(x)\tan(\frac{1}{2}\gamma\zeta)+2\tan(\frac{1}{2}\gamma\zeta)}{1-\tan^2(x)\tan^2(\frac{1}{2}\gamma\zeta)}\Bigr)+\frac{\tan^2(x)-\tan^2(\frac{1}{2}\gamma\zeta)}{1-\tan^2(x)\tan^2(\frac{1}{2}\gamma\zeta)}}{\tanh^2\Bigl(\frac{\zeta}{2}\Bigr)+i\tanh\Bigl(\frac{\zeta}{2}\Bigr)\Bigl(\frac{2\tan^2(x)\tan(\frac{1}{2}\gamma\zeta)+2\tan(\frac{1}{2}\gamma\zeta)}{1-\tan^2(x)\tan^2(\frac{1}{2}\gamma\zeta)}\Bigr)+\frac{\tan^2(x)-\tan^2(\frac{1}{2}\gamma\zeta)}{1-\tan^2(x)\tan^2(\frac{1}{2}\gamma\zeta)}}\nonumber\\
=&\frac{\tanh^2\Bigl(\frac{\zeta}{2}\Bigr)\Bigl(1-\tan^2(x)\tan^2\Bigl(\frac{1}{2}\gamma\zeta\Bigr)\Bigr)+\tan^2(x)-\tan^2\Bigl(\frac{1}{2}\gamma\zeta\Bigr)-iT(x,\zeta,\gamma)}{\tanh^2\Bigl(\frac{\zeta}{2}\Bigr)\Bigl(1-\tan^2(x)\tan^2\Bigl(\frac{1}{2}\gamma\zeta\Bigr)\Bigr)+\tan^2(x)-\tan^2\Bigl(\frac{1}{2}\gamma\zeta\Bigr)+iT(x,\zeta,\gamma)}\nonumber\\
=&\frac{\tanh^2\Bigl(\frac{\zeta}{2}\Bigr)-2i\tanh\Bigl(\frac{\zeta}{2}\Bigr)\tan\Bigl(\frac{1}{2}\gamma\zeta\Bigr)-\tan^2\Bigl(\frac{1}{2}\gamma\zeta\Bigr)+\tan^{2}(x)S(\zeta,\gamma)_{-}}{\tanh^2\Bigl(\frac{\zeta}{2}\Bigr)+2i\tanh\Bigl(\frac{\zeta}{2}\Bigr)\tan\Bigl(\frac{1}{2}\gamma\zeta\Bigr)-\tan^2\Bigl(\frac{1}{2}\gamma\zeta\Bigr)+\tan^{2}(x)S(\zeta,\gamma)_{+}}
\label{tan_Im2}
\end{eqnarray}
where 
\begin{eqnarray}
S(\zeta,\gamma)_{\pm}:=
-\tanh^2\Bigl(\frac{\zeta}{2}\Bigr)\tan^2\Bigl(\frac{1}{2}\gamma\zeta\Bigr)\pm2i\tanh\Bigl(\frac{\zeta}{2}\Bigr)\tan\Bigl(\frac{1}{2}\gamma\zeta\Bigr)+1\\
T(x,\zeta,\gamma):=
\tanh\Bigl(\frac{\zeta}{2}\Bigr)\Bigl(2\tan^2(x)\tan\Bigl(\frac{1}{2}\gamma\zeta\Bigr)+2\tan\Bigl(\frac{1}{2}\gamma\zeta\Bigr)\Bigr).
\end{eqnarray}

\subsection{The expression of $\tan^{2}(x)$ using $\gamma$}\label{sec:repre_tan^2}

We now derive the expression of $\tan^{2}(x)$ as a function of $\gamma$ and $n$. 
We define $X$ by $X\equiv\tan^{2}{x}$. Relation: $ (\ref{tan_Im2})=\exp\biggl(\frac{2\pi in}{N}\biggr)\biggl(\frac{(\tanh(\zeta)+i\tan(-\gamma\zeta))^2}{(\tanh(\zeta)-i\tan(-\gamma\zeta))^2}\biggr)^{\frac{1}{N}}$ is expressed as

\begin{eqnarray}
\frac{A+BX}{\tilde{A}+\tilde{B}X}=C\label{eq:AA'BB'Ceq}
\end{eqnarray}
where $A$, $B$, $\tilde{A}$, $\tilde{B}$ and, $C$ are
\begin{eqnarray}
A&=\tanh^2\Bigl(\frac{\zeta}
{2}\Bigr)-2i\tanh\Bigl(\frac{\zeta}{2}\Bigr)\tan\Bigl(\frac{1}{2}\gamma\zeta\Bigr)-\tan^2\Bigl(\frac{1}{2}\gamma\zeta\Bigr)\\
B&=-\tanh^2\Bigl(\frac{\zeta}{2}\Bigr)\tan^2\Bigl(\frac{1}{2}\gamma\zeta\Bigr)-2i\tanh\Bigl(\frac{\zeta}{2}\Bigr)\tan\Bigl(\frac{1}{2}\gamma\zeta\Bigr)+1\\
\tilde{A}&=\tanh^2\Bigl(\frac{\zeta}{2}\Bigr)+2i\tanh\Bigl(\frac{\zeta}{2}\Bigr)\tan\Bigl(\frac{1}{2}\gamma\zeta\Bigr)-\tan^2\Bigl(\frac{1}{2}\gamma\zeta\Bigr)\\
\tilde{B}&=-\tanh^2\Bigl(\frac{\zeta}{2}\Bigr)\tan^2\Bigl(\frac{1}{2}\gamma\zeta\Bigr)+2i\tanh\Bigl(\frac{\zeta}{2}\Bigr)\tan\Bigl(\frac{1}{2}\gamma\zeta\Bigr)+1\\
C&=\exp\biggl(\frac{2\pi in}{N}\biggr)\biggl(\frac{(\tanh(\zeta)+i\tan(-\gamma\zeta))^2}{(\tanh(\zeta)-i\tan(-\gamma\zeta))^2}\biggr)^{\frac{1}{N}}.
\end{eqnarray}
From the relation (\ref{eq:AA'BB'Ceq}), X is expressed by
\begin{eqnarray}
X=\frac{C\tilde{A}-A}{B-C\tilde{B}}.
\end{eqnarray}
Thus, we obtain
\begin{eqnarray}
&\tan^2(x)=\frac{\exp\biggl(\frac{2\pi in}{N}\biggr)\biggl(\frac{(\tanh(\zeta)+i\tan(-\gamma\zeta))^2}{(\tanh(\zeta)-i\tan(-\gamma\zeta))^2}\biggr)^{\frac{1}{N}}E(\zeta, \frac{1}{2}\gamma\zeta)_{+}-E(\zeta, \frac{1}{2}\gamma\zeta)_{-}}{D(\zeta, \frac{1}{2}\gamma\zeta)_{-}-\exp\biggl(\frac{2\pi in}{N}\biggr)\biggl(\frac{(\tanh(\zeta)+i\tan(-\gamma\zeta))^2}{(\tanh(\zeta)-i\tan(-\gamma\zeta))^2}\biggr)^{\frac{1}{N}}D(\zeta, \frac{1}{2}\gamma\zeta)_{+}}\nonumber\\
\end{eqnarray}
where 
\begin{eqnarray}
D(\zeta, \phi)_{\pm}\equiv-\tanh^2\Bigl(\frac{\zeta}{2}\Bigr)\tan^2(\phi)\pm2i\tanh\Bigl(\frac{\zeta}{2}\Bigr)\tan(\phi)+1\\
E(\zeta, \phi)_{\pm}\equiv\tanh^2\Bigl(\frac{\zeta}{2}\Bigr)\pm2i\tanh\Bigl(\frac{\zeta}{2}\Bigr)\tan(\phi)-\tan^2(\phi).
\end{eqnarray}
$\phi$ is defined by
\begin{eqnarray}
\phi\equiv\frac{1}{2}\gamma\zeta\ \biggl(\mbox{i.e.}\ \gamma=\frac{2\phi}{\zeta}\biggr).\label{eq:definition_of_phi}
\end{eqnarray}
Moreover, we remark 
\begin{eqnarray}
\Biggl(\frac{\tanh(\zeta)+i\tan(-\gamma\zeta)}{\tanh(\zeta)-i\tan(-\gamma\zeta)}\Biggr)^2&=\Biggl(\frac{\tanh(\zeta)-i\tan(\gamma\zeta)}{\tanh(\zeta)+i\tan(\gamma\zeta)}\Biggr)^2\nonumber\\
&=\Biggl(\frac{\tanh(\zeta)-i\frac{2\tan(\frac{1}{2}\gamma\zeta)}{1-\tan^2(\frac{1}{2}\gamma\zeta)}}{\tanh(\zeta)+i\frac{2\tan(\frac{1}{2}\gamma\zeta)}{1-\tan^2(\frac{1}{2}\gamma\zeta)}}\Biggr)^2\nonumber\\
&=\Biggl(\frac{\tanh(\zeta)\Bigl(1-\tan^2\Bigl(\frac{1}{2}\gamma\zeta\Bigr)\Bigr)-i2\tan\Bigl(\frac{1}{2}\gamma\zeta\Bigr)}{\tanh(\zeta)\Bigl(1-\tan^2\Bigl(\frac{1}{2}\gamma\zeta\Bigr)\Bigr)+i2\tan\Bigl(\frac{1}{2}\gamma\zeta\Bigr)}\Biggr)^2\nonumber\\
&=\biggl(L(\zeta, \frac{1}{2}\gamma\zeta)\biggr)^{2}
\end{eqnarray}
where
\begin{eqnarray}
L(\zeta, \phi)\equiv\biggl(\frac{\tanh(\zeta)(1-\tan^2(\phi))-i2\tan(\phi)}{\tanh(\zeta)(1-\tan^2(\phi))+i2\tan(\phi)}\biggr).
\end{eqnarray}
Therefore, we obtain $\tan^{2}(x)$ as follows.
\begin{eqnarray}
\hspace*{0mm}
\tan^2(x)&=\frac{\exp\biggl(\frac{2\pi in}{N}\biggr)\biggl(\frac{(\tanh(\zeta)+i\tan(-\gamma\zeta))^2}{(\tanh(\zeta)-i\tan(-\gamma\zeta))^2}\biggr)^{\frac{1}{N}}E(\zeta, \frac{1}{2}\gamma\zeta)_{+}-E(\zeta, \frac{1}{2}\gamma\zeta)_{-}}{D(\zeta, \frac{1}{2}\gamma\zeta)_{-}-\exp\biggl(\frac{2\pi in}{N}\biggr)\biggl(\frac{(\tanh(\zeta)+i\tan(-\gamma\zeta))^2}{(\tanh(\zeta)-i\tan(-\gamma\zeta))^2}\biggr)^{\frac{1}{N}}D(\zeta, \frac{1}{2}\gamma\zeta)_{+}}\nonumber\\
&=\frac{\exp\biggl(\frac{2\pi in}{N}\biggr)\biggl(\biggl(L(\zeta, \frac{1}{2}\gamma\zeta)\biggr)^2\biggr)^{\frac{1}{N}}E(\zeta, \frac{1}{2}\gamma\zeta)_{+}-E(\zeta, \frac{1}{2}\gamma\zeta)_{-}}{D(\zeta, \frac{1}{2}\gamma\zeta)_{-}-\exp\biggl(\frac{2\pi in}{N}\biggr)\biggl(\biggl(L(\zeta, \frac{1}{2}\gamma\zeta)\biggr)^2\biggr)^{\frac{1}{N}}D(\zeta, \frac{1}{2}\gamma\zeta)_{+}}\nonumber\\
&=\frac{\exp\biggl(\frac{2\pi in}{N}\biggr)\biggl(\biggl(L(\zeta, \phi)\biggr)^2\biggr)^{\frac{1}{N}}E(\zeta, \phi)_{+}-E(\zeta, \phi)_{-}}{D(\zeta, \phi)_{-}-\exp\biggl(\frac{2\pi in}{N}\biggr)\biggl(\biggl(L(\zeta, \phi)\biggr)^2\biggr)^{\frac{1}{N}}D(\zeta, \phi)_{+}}.
\label{eq:real1}
\end{eqnarray}

Thus, we have derived a systematic method for numerically deriving the solutions of the Bethe ansatz equations (\ref{eq:BAE11}) and (\ref{eq:BAE22}).
If we use eq. (\ref{eq:real1}) with a fixed integer $n$, the first Bethe ansatz equation  (\ref{eq:BAE11}) becomes the equation for only one variable $\phi$. Thus, if we fix the Bethe quantum number $J_{1}$, we numerically derive the value of $\phi$ corresponding to $J_1$ from eqs. (\ref{eq:real1}) and (\ref{eq:real_counting_function_XXZ}). 
We readily evaluate the deviation $\gamma$ of the Bethe solution by eq. (\ref{eq:definition_of_phi}). 

Furthermore, from eq. (\ref{eq:real1}), if we fix $n$, 
we can evaluate the center of the Bethe ansatz equations $x$. 
Here the integer $n$ corresponds to the difference between the Bethe quantum numbers $J_{1}$ and $J_{2}$. We shall explain it in \ref{sec:massive_real_quantum}.
In particular, (\ref{eq:real1}) is useful for deriving the conditions for the occurrence of collapse and extra two-string solution when $n=0$.
This derivation is discussed in Section \ref{sec:collapse_extra}.

\section{Revisit of collapsed solutions and the emergence of extra two-string solutions}\label{sec:collapse_extra}

In this section, we derive the condition of the collapse and emergence of extra two-string solutions from the real solutions.
The number of real solutions is consistent with that of complex solutions.

\subsection{Limit of the counting function $W(\phi)$ as $\phi\rightarrow0$}

In the complex solution, it is known that when the inequality (\ref{eq:condition_of_extra-two-string_form_complex_solution}) is satisfied, the extra-two-string solutions emerge at the Bethe quantum numbers $ J_{1}=J_{2}=\frac{N-1}{2}$.\cite{ISD}
On the other hand when the inequality(\ref{eq:condition_of_N_collapse_form_complex_solution}) is satisfied,
the collapse of $m$ two-string solutions occurs.\cite{ISD}

In this section, we analyze the real solutions in the two down-spin sector and obtain the condition of the emergence of extra-two-string solutions and the occurrence of collapse.

We define $W(\phi)$ as $\phi=\frac{\gamma\zeta}{2}$, fixing $\zeta$ and substituting x in (\ref{eq:real1}) for $W(x,\gamma,\zeta)$.
When two rapidities of real solutions are close to complex solutions, the deviation $\phi$ is close to $0$ (see Figure \ref{fig:move_on complex_plain}). Thus we calculate $\tan^{2}{(x)(\phi)}$ (\ref{eq:real1}) in the limit of $\phi\rightarrow0$ with $\phi>0$.
We expand $\tan^{2}x(\phi)$ with respect to $\phi$ as follows.

\begin{eqnarray}
\tan^{2}x(\phi)=\frac{2\coth(\zeta)\tanh^{2}(\frac{\zeta}{2})-N\tanh(\frac{\zeta}{2})}{N\tanh(\frac{\zeta}{2})-2\coth(\zeta)}\nonumber\\
+\frac{2}{3N\bigl(N\tanh(\frac{\zeta}{2})-2\coth(\zeta)\bigr)^{2}}
\biggl(6N\coth^{2}(\zeta)\nonumber\\
-6N^{2}\coth(\zeta)\tanh(\frac{\zeta}{2})-16\coth^{3}(\zeta)\tanh(\frac{\zeta}{2})\nonumber\\
+4N^{2}\coth(\zeta)^{3}\tanh(\frac{\zeta}{2})+6N^{2}\coth(\zeta)\tanh^{3}(\frac{\zeta}{2})\nonumber\\
+16\coth^{3}(\zeta)\tanh^{3}(\frac{\zeta}{2})-4N^{2}\coth^{3}(\zeta)\tanh^{3}(\frac{\zeta}{2})\nonumber\\
-6N\coth^{2}(\zeta)\tanh^{4}(\frac{\zeta}{2})
\biggr)\phi^{2}+O(\phi^{4})
\end{eqnarray}
It is easy to show that
\begin{equation}
\tan^{2}{x(\phi)}\rightarrow\frac{2\coth(\zeta)\tanh^{2}(\frac{\zeta}{2})-N\tanh(\frac{\zeta}{2})}{N\tanh(\frac{\zeta}{2})-2\coth(\zeta)}\ \ \ (\phi\rightarrow0).
\end{equation}
Thus, we obtain the Bethe ansatz equation (\ref{eq:real_counting_function_XXZ}) in the limit as $\phi$ approaches $0$ as follows.

\begin{eqnarray}
\frac{2\pi}{N}J_{1}&=&2\pi W(x,\phi=0,\zeta)\nonumber\\
&=&2\tan^{-1}\biggl(\frac{\tan(x)}{\tanh(\zeta/2)}\biggr)\nonumber\\
&=&2\tan^{-1}\biggl(\sqrt{\frac{N-(1+t^{2})}{1-(N-1)t^{2}}}\biggr)
\end{eqnarray}
where $ t=\tanh\bigl(\frac{\zeta}{2}\bigr)$.

\begin{figure}[htbp]
\begin{center}
 \includegraphics[clip,width=15.0cm]{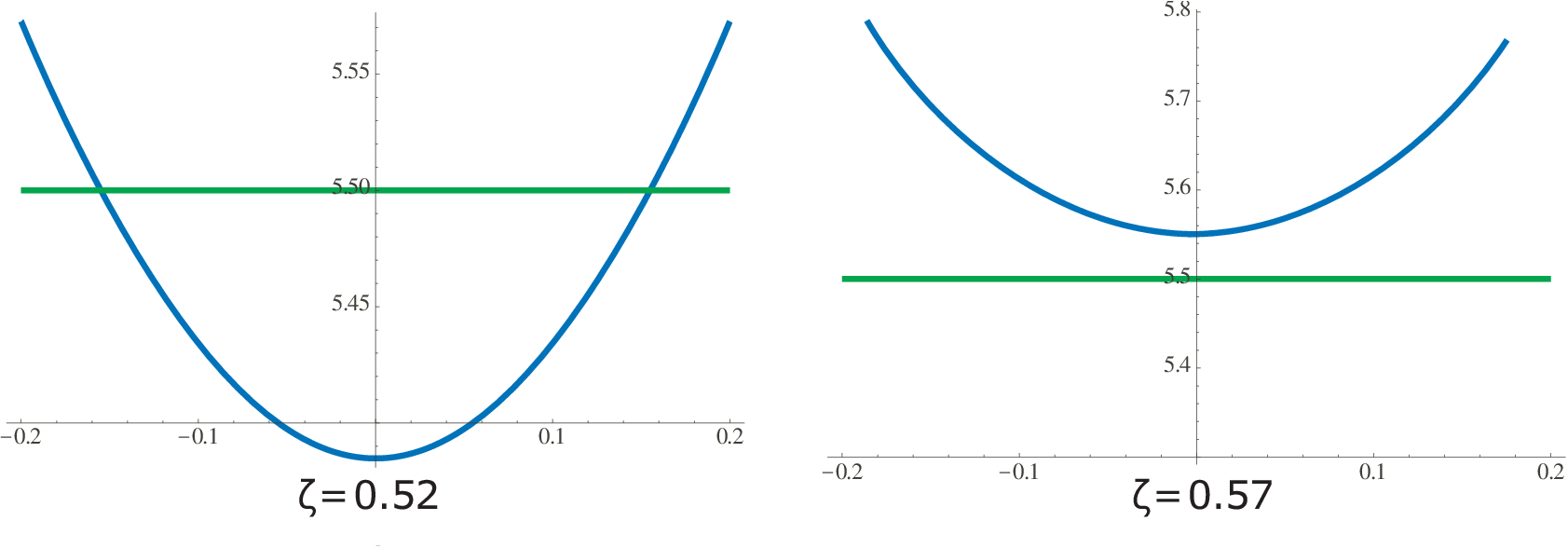}
\end{center}
\caption{Graph of counting function. The blue graph is $NW(\phi)$ against $\phi$ where $N$ is the site number and $W(\phi)$ is the counting function. The green graph is $ \frac{N-1}{2}$. In this case, the site number is $N=12$, the anisotropic parameter is $\zeta=0.52$ in the left figure, and the anisotropic parameter is $\zeta=0.57$ in the right figure. When the crossing points of the blue graph and the green graph exist, the real solution corresponding to the Bethe quantum number $J_{1}=J_{2}=\frac{N-1}{2}$ exists. The left figure has the real solutions corresponding to the Bethe quantum number $ J_{1}=\frac{N-1}{2}$. On the other hand, the right figure does not have the real solution corresponding to the Bethe quantum number $ J_{1}=\frac{N-1}{2}$, but the complex solution corresponding to the Bethe quantum number $ J_{1}=\frac{N-1}{2}$.} 
\label{fig:counting_function_for_the_real_and_the_complex_solution}
\end{figure}

Figure \ref{fig:counting_function_for_the_real_and_the_complex_solution} illustrates the counting function for the site number $N = 12$ and the anisotropic parameter with $\zeta = 0.52$ and $0.57$, respectively.

\subsection{Condition of the collapses and that of the emergence of an extra two-string solution}

From the previous section, we obtain the conditions that an extra two-string solution emerges in the chain of $N$ site given by
\begin{eqnarray}
\frac{N-1}{2}<\frac{N}{\pi}\tan^{-1}\biggl(\sqrt{\frac{N-(1+t^{2})}{1-(N-1)t^{2}}}\biggr).\label{eq:extra_two_string_condition}
\end{eqnarray}
On the other hand, we obtain the conditions that the collapse of $m$ two-string solutions occurs in the chain of $N$ site for $m=1,2,\cdots$ are given by
\begin{eqnarray}
\frac{N-(3+2m)}{2}<\frac{N}{\pi}\tan^{-1}\biggl(\sqrt{\frac{N-(1+t^{2})}{1-(N-1)t^{2}}}\biggr)<\frac{N-(1+2m)}{2}.\label{eq:collapse_condition}
\end{eqnarray}

\begin{figure}[htbp]
\begin{center}
 \includegraphics[clip,width=15.0cm]{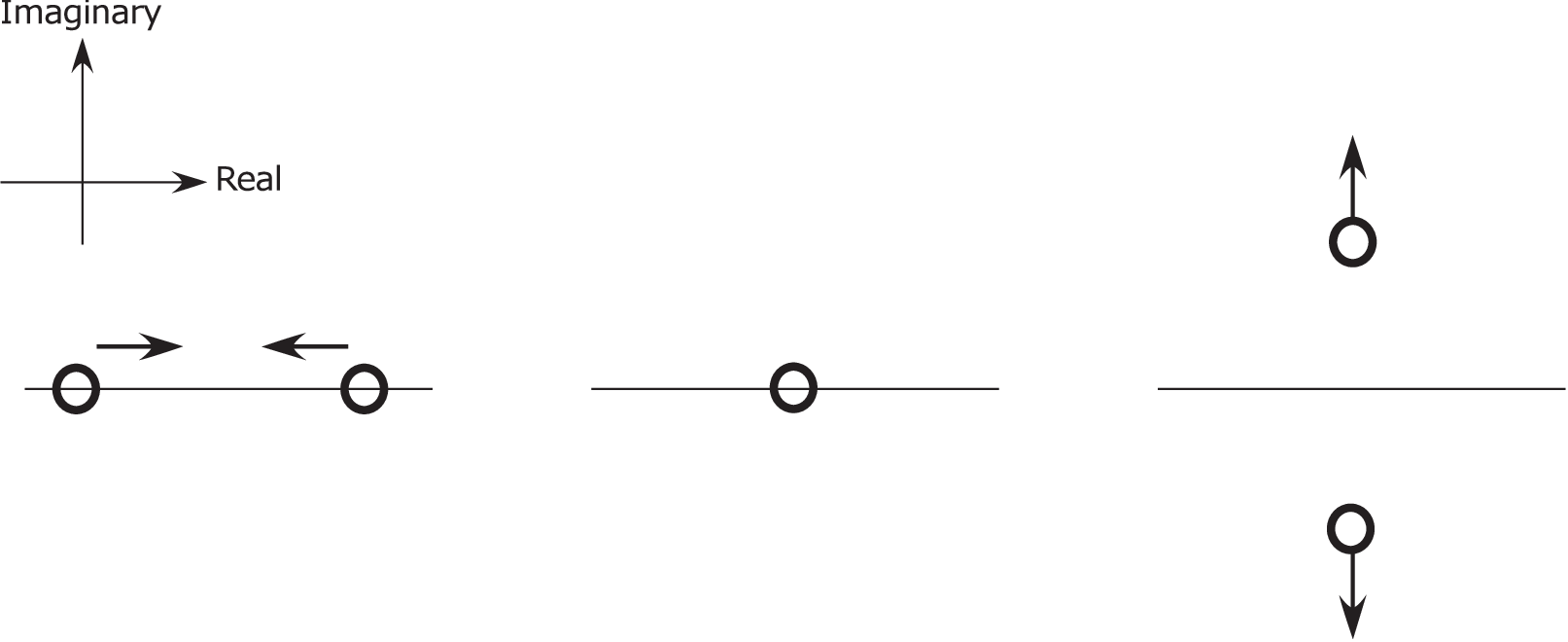}
\end{center}
\caption{Pair of the rapidities of a Bethe solution in the complex plane.
We assign the imaginary part of the Bethe solutions in the vertical axis and the real part of the Bethe solutions in the horizontal axis.
The left figure depicts the real solution. The Bethe solution moves with respect to $\zeta$  on the real axis.
If the anisotropy parameter $\zeta$ is large, the two rapidities approach each other on the real axis.
The right figure depicts the complex solution. The Bethe solution moves with respect to $\zeta$ on the complex plain.
If the anisotropy parameter $\zeta$ is large, the two rapidities move away from each other on the imaginary axis.
The central figure depicts the critical situation between the real solution and the complex solution. In this case, the deviation of two real Bethe solutions $\phi$ is zero. 
}
\label{fig:move_on complex_plain}
\end{figure}

This result is consistent with that in the complex solution\cite{ISD}.
Figure \ref{fig:move_on complex_plain} illustrates the behavior of the Bethe solution corresponding to the Bethe quantum number $J_{1}=J_{2}=\frac{N-1}{2}$.

\section{The divergence of some of the Bethe solutions in the XXX limit}
\label{sec:divergence}

There exist $_{N}C_{2}-{}_{N}C_{1}$ solutions for the XXX chain in the two down-spin sector. 
However, it is known that there exist $_{N}C_{2}$ solutions for the massive XXZ chain in the two down-spin sector.
In this section, we prove the divergence of the infinite Bethe solutions of the massive XXZ chain in the XXX limit.

\subsection{The value of the height function $h(\zeta, \lambda_{1})$ in $J_{1}=\frac{N-1}{2}$ and $\lambda_{1}=\frac{\pi}{4}$}\label{sec:h_limit_lambda1_pi4}

Let us assume the first Bethe quantum number $J_{1}=\frac{N-1}{2}$ and the Bethe solutions $\lambda_{1},\ \lambda_{2}>0$.
We define $\bar{\lambda}_{1}$, $\bar{\lambda}_{2}$ by $\bar{\lambda}_{1}=\lambda_{1}/\zeta$, $\bar{\lambda}_{2}=\lambda_{2}/\zeta$.
In this case, the solution of the Bethe ansatz equation $\lambda_{1},\lambda_{2}$ satisfy the inequalities $\bigl(-\frac{\pi}{2}<\bigr)\bar{\lambda}_{1}\zeta-\bar{\lambda}_{2}\zeta<\frac{\pi}{2}$. 
We prove these inequalities in the \ref{sec:app_ineq}.

If $0<\zeta\ll1$ and $\lambda_{1}=\frac{\pi}{4}$, we have
\begin{eqnarray}
\lambda_{2}\Bigl(\lambda_{1}=\frac{\pi}{4}\Bigr)&=&\tan^{-1}\biggl(-\frac{\frac{\tanh(\zeta)}{\tan(N\tan^{-1}(1/\tanh(\zeta/2)))}+1}{\frac{\tanh(\zeta)}{\tan(N\tan^{-1}(1/\tanh(\zeta/2)))}-1}\biggr)\nonumber\\
&=&\tan^{-1}\Bigl(\frac{1+t}{1-t}\Bigr)\label{eq:lambda_2_lambda_1_pi/4}
\end{eqnarray}
where $t\equiv\frac{\tanh(\zeta)}{\tan(N\tan^{-1}(1/\tanh(\zeta/2)))}$.
From the relation (\ref{eq:lambda_2_lambda_1_pi/4}) and the inequality $t>0$, we estimate the height function $h(\zeta, \lambda_{1}=\frac{\pi}{4})$ as follows. 
\begin{eqnarray}
h\Bigl(\zeta,\lambda_{1}=\frac{\pi}{4}\Bigr)&=&\frac{N}{\pi}\tan^{-1}\Bigl(\frac{1}{\tanh\zeta/2}\frac{1+t}{1-t}\Bigr)-\frac{1}{\pi}\tan^{-1}\Bigl(\frac{\tan(\lambda_{2}-\frac{\pi}{4})}{\tanh\zeta}\Bigr)\nonumber\\
&=&\frac{N}{\pi}\tan^{-1}\Bigl(\frac{1}{\tanh\zeta/2}\frac{1+t}{1-t}\Bigr)-\frac{1}{\pi}\tan^{-1}\Bigl(\frac{t}{\tanh\zeta}\Bigr)
\end{eqnarray}
Thus,
\begin{eqnarray}
h\Bigl(\zeta,\lambda_{1}=\frac{\pi}{4}\Bigr)>\frac{N-1}{2}\label{eq:h_limit_pi4}
\end{eqnarray}

\subsection{The continuity of the height function $h(\bar{\lambda}_{1}\zeta)$ in the interval $\displaystyle \frac{\pi}{4}<\bar{\lambda}_{1}\zeta<\frac{\pi}{2}$}\label{sec:The_continuity_of_J_{2}(lamb_{1})}

In this subsection, we prove the continuity of the height function $h(\bar{\lambda}_{1}\zeta)$ in the interval $\frac{\pi}{4}<\bar{\lambda}_{1}\zeta<\frac{\pi}{2}$. 
At first, we show the continuity of $\bar{\lambda}_{1}\zeta$.
Both the numerator and the denominator of $\tan(\bar{\lambda}_{2}\zeta)$ is continuous. 
Thus, if the sign of the denominator of $\tan(\bar{\lambda}_{2}\zeta)$:
\begin{eqnarray}
\tan(\bar{\lambda}_{1}\zeta)+\tanh(\zeta)\tan\biggl(N\tan^{-1}\Bigl(\frac{\tan(\bar{\lambda}_{1}\zeta)}{\tanh(\frac{\zeta}{2})}\Bigr)-\pi\frac{N-1}{2}\biggr)\label{eq:denominator_conti}
\end{eqnarray}
does not change, we obtain the continuity of the function $\tan(\bar{\lambda}_{2}\zeta)$ of $\bar{\lambda}_{1}\zeta$. 
If $\zeta$ is sufficiently small, we show
\begin{eqnarray}
&&N\tan^{-1}\biggl(\frac{\tan(\bar{\lambda}_{1}\zeta)}{\tanh(\frac{\zeta}{2})}\biggr)-\pi\frac{N-1}{2}\nonumber\\
>&&N\tan^{-1}\biggl(\frac{1}{\tanh(\frac{\zeta}{2})}\biggr)-\pi\frac{N-1}{2}\nonumber\\
>&&0.
\end{eqnarray}
Thus, we show (\ref{eq:denominator_conti}) is positive and $\tan(\bar{\lambda}_{1}\zeta)$ is continuous.
From the continuity of $\tan(\bar{\lambda}_{1}\zeta)$ and $\tan(\bar{\lambda}_{2}\zeta)$ and $\bar{\lambda}_{1}, \bar{\lambda}_{2}>0$, we show the continuity of $h(\bar{\lambda}_{1}\zeta)$ in the interval  $\frac{\pi}{4}<\bar{\lambda}_{1}\zeta<\frac{\pi}{2}$.

\subsection{The divergence of the Bethe solutions}
\label{sec:The_divergence_of_the_Bethe_solutions}
From the equations (\ref{eq:height_fuc_pi/2lim_odd}), (\ref{eq:height_fuc_pi/2lim_even}) the inequality (\ref{eq:h_limit_pi4}), and continuity of the height function $h(\bar{\lambda}_{1}\zeta)$ in the interval $\frac{\pi}{4}<\bar{\lambda}_{1}\zeta<\frac{\pi}{2}$ in the subsection \ref{sec:The_continuity_of_J_{2}(lamb_{1})},
when the Bethe quantum numbers
\begin{eqnarray}
J_{1}=\frac{N-1}{2}, J_{2}=\frac{1}{2},\cdots,\frac{N-1}{2}\label{eq:bqn_infinite_positive}
\end{eqnarray}
the solution of the Bethe ansatz equation is satisfied with the condition:
\begin{eqnarray}
\frac{\pi}{4}\leq\bar{\lambda}_{1}\zeta<\frac{\pi}{2}\label{eq:lambda1_zeta_inequality}
\end{eqnarray}
We divide every side of the inequality (\ref{eq:lambda1_zeta_inequality}) by $\zeta$ as follows. 
\begin{eqnarray}
\frac{\pi}{4\zeta}\leq\bar{\lambda}_{1}<\frac{\pi}{2\zeta}
\end{eqnarray}
When we send $\zeta$ to zero, we show the reduced rapidities $\bar{\lambda}_{1}$ diverge to infinity: $\bar{\lambda}_{1}\rightarrow\infty$.
Similarly, when the Bethe quantum numbers $J_{1}$ and $J_{2}$ satisfy the following conditions\footnote{When the site number $N$ is odd, the Bethe quantum numbers are given by $J_{1}=\frac{N-1}{2}$, $J_{2}=0,1,\cdots,\frac{N-1}{2}$.\\ We remark $(\frac{N-1}{2},0)$ and $(-\frac{N-1}{2},0)$ is same solution. Therefore, the number of solutions is $(\frac{N-1}{2}+1)+(\frac{N-1}{2}-1)-1=N(={}_{N}C_{1})$}:
\begin{eqnarray}
J_{1}=-\frac{N-1}{2},\ J_{2}=-\frac{1}{2},\cdots,-\frac{N-1}{2},\label{eq:bqn_infinite_negative}
\end{eqnarray}
the Bethe solutions are satisfied with the following conditions:
\begin{eqnarray}
-\frac{\pi}{4}\geq\bar{\lambda}_{1}\zeta>-\frac{\pi}{2}\label{eq:minus_lambda1_zeta_inequality}
\end{eqnarray}
We divide every side of the inequality (\ref{eq:minus_lambda1_zeta_inequality}) by $\zeta$ as follows. 
\begin{eqnarray}
-\frac{\pi}{4\zeta}\geq\bar{\lambda}_{1}>-\frac{\pi}{2\zeta}.
\end{eqnarray}
Therefore, when we send $\zeta$ to zero, we show the reduced rapidities $\bar{\lambda}_{1}$ diverge to minus infinity: $\bar{\lambda}_{1}\rightarrow-\infty$.

\section{The completeness of the massive XXZ spin chain in the two down-spin sector}
\label{sec:completeness}

\subsection{The proof of the completeness}

It is known that in the XXX spin chain for the two down-spin sector the number of the Bethe solutions is ${}_{N}C_{2}-{}_{N}C_{1}$\cite{DG1}.
The number of the finite Bethe solutions is eqaul to the number of the Bethe solutions in the XXX spin chain for the two down-spin sector.
On the other hand, the number of the infinite Bethe solutions corresponding to both  positive and negative quantum numbers is  
\begin{eqnarray}
\frac{N}{2}+\frac{N}{2}={}_{N}C_{1}.
\end{eqnarray}
Similarly, when the site number $N$ is odd, the number of the Bethe quantum number is ${}_{N}C_{1}$.
Therefore, the number of the total Bethe quantum numbers is
\begin{eqnarray}
{}_{N}C_{2}-{}_{N}C_{1}+{}_{N}C_{1}={}_{N}C_{2}.
\end{eqnarray}
The number is consistent with that of other approach of the completeness of the XXZ spin chain for the two down-spin sector, which is proven by Koma and Ezawa\cite{KE}.

\subsection{The list of the Bethe quantum numbers}
We obtain every Bethe quantum number of the Bethe ansatz equation for the massive XXZ spin chain in the two down-spin sector.
In this subsection, we list the every Bethe quantum number including the complex solutions. 
We remark that extra two-string solutions and the collapse of two-string solutions exist.

\begin{figure}[htbp]
\begin{center}
 \includegraphics[clip,width=16cm]{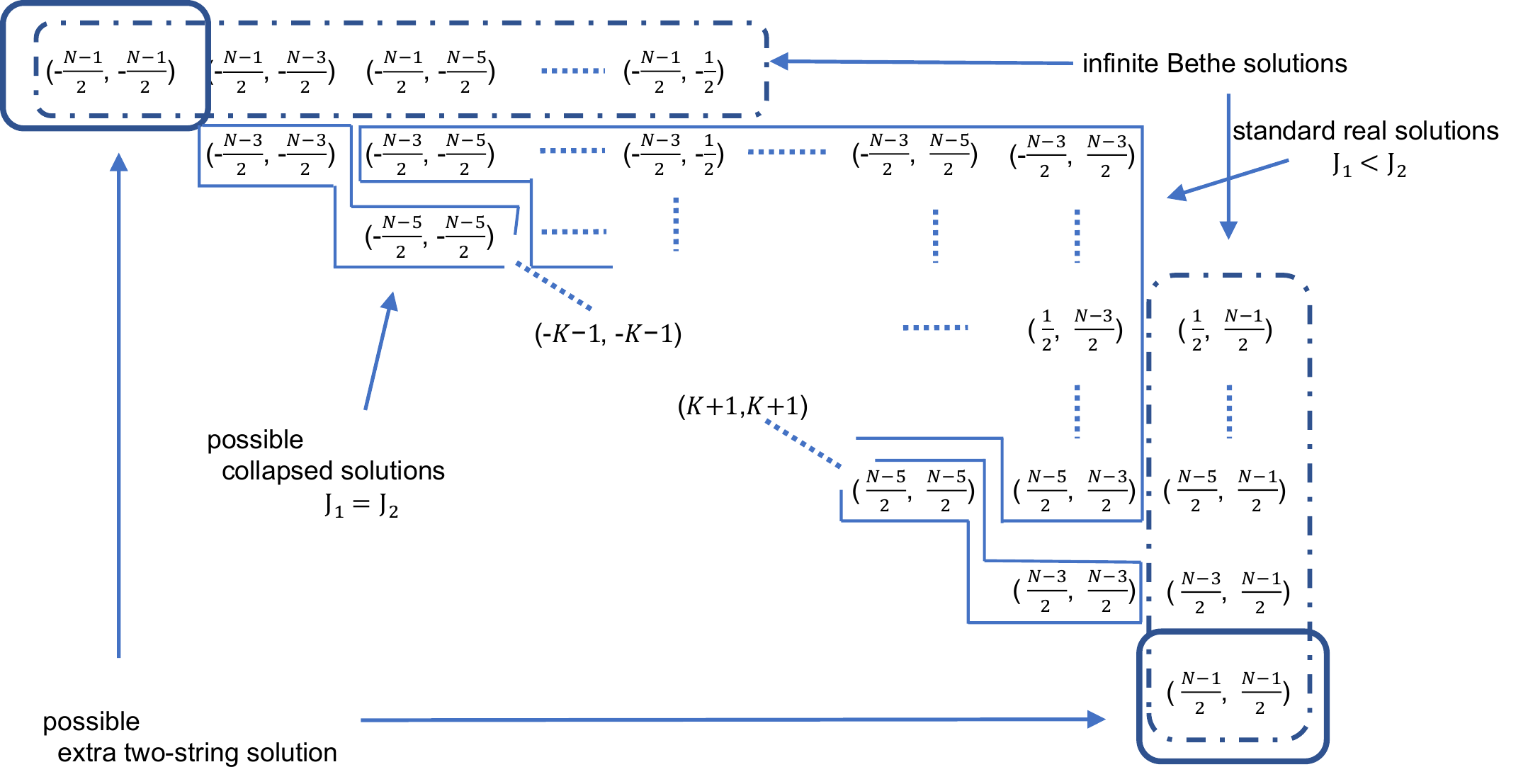}
\end{center}
 \caption{
List of the Bethe quantum numbers where the $N$ is even.
 The standard real solution (i. e. $-\frac{N-1}{2}<J_{1}<J_{2}<\frac{N-1}{2}$) is corresponding to $J_{1}<J_{2}$ in the figure.
 This correspond to (\ref{enum:normal_real_solution})
 Infinite Bethe solution correspond to (I\hspace{-.1em}I\hspace{-.1em}I).
 The Bethe quantum number $(\frac{N-3}{2}, \frac{N-3}{2})\cdots (-\frac{N-3}{2}, -\frac{N-3}{2})$ correspond to the collapsed solution.
 It correspond to (\ref{enum:m=0_real_solution})
  The Bethe quantum number $(\frac{N-1}{2}, \frac{N-1}{2})$ and $(-\frac{N-1}{2}, -\frac{N-1}{2})$ correspond to the extra two-string solution.
  It correspond to (\ref{enum:narrow_pair}) or (\ref{enum:enum:inf_solution})
}
\label{fig:bqe_list} 
\end{figure}

\begin{enumerate}[(I)]
\item The complex solutions for the Bethe ansatz equation\\ 
In this case, the Bethe quantum numbers are given by \cite{DG1} and \cite{ISD}.
We divide the regime of $\zeta$ and $N$ into two.
We call such a regime of $\zeta$ and $N$ satisfying $\tanh^{2}(\frac{\zeta}{2})\geq\frac{1}{N-1}$ for $\zeta>0$ the stable regime.
On the other hand, we call such a regime of $\zeta$ and $N$ satisfying $\tanh^{2}(\frac{\zeta}{2})\leq\frac{1}{N-1}$ for $\zeta>0$ the unstable regime. 
When the absolute value of the imaginary part of the Bethe solution is more than $\frac{\zeta}{2}$, we call the pair of the Bethe solution the wide pair.
When the absolute value of the imaginary part of the Bethe solution is less than $\frac{\zeta}{2}$, we call the pair of the Bethe solution the narrow pair.
In the stable regime, we combine (\si) and (\sii), while in the unstable regime, we combine (\si) and (\siii).

\begin{enumerate}[(i)]\label{enum:complex}
\item The wide pair\\
The Bethe quantum number $J_{1}$ satisfy
\begin{eqnarray}
\frac{N}{4}-\frac{1}{2}<J_{1}<\frac{N-1}{2}\ \mbox{for}\ \biggl(\tan(x)>0\biggr),\label{eq:wide_pair_plus}\\
-\frac{N+1}{2}<J_{1}<-\frac{N}{4}-\frac{1}{2}\ \mbox{for}\ \biggl(\tan(x)<0\biggr).\label{eq:wide_pair_minus}
\end{eqnarray}
In this case another Bethe quantum number $J_{2}$ is equal to $J_{1}+1$: $J_{2}=J_{1}+1$ 
\item The narrow pair(stable regime)\label{enum:narrow_pair}\\
\begin{eqnarray}
\frac{N}{4}<J_{1}<\frac{N}{2}\ \mbox{for}\ \biggl(\tan(x)>0\biggr),\\
-\frac{N}{2}<J_{1}<-\frac{N}{4}\ \mbox{for}\ \biggl(\tan(x)<0\biggr).
\end{eqnarray}
In this case another Bethe quantum number $J_{2}$ is equal to $J_{1}$: $J_{2}=J_{1}$ 
\item The narrow pair(unstable regime)\label{enum:unstable}\\
\begin{eqnarray}
\frac{N}{4}< J_{1}<\frac{N}{\pi}\tan^{-1}\biggl(\sqrt{\frac{N-(1+\tanh^{2}(\zeta/2))}{1-(N-1)\tanh^{2}(\zeta/2)}}\biggr)\nonumber\\
\ \ \ \ \ \ \ \ \ \ \ \ \ \ \ \ \ \ \ \ \ \ \ \ \ \ \ \ \ \ \ \ \ \ \ \ \ \ \ \ \ \ \ \ \mbox{for}\ \biggl(\tan(x)>0\biggr),\label{eq:unstable_plus}\\
-\frac{N}{4}> J_{1}>\frac{N}{\pi}\tan^{-1}\biggl(\sqrt{\frac{N-(1+\tanh^{2}(\zeta/2))}{1-(N-1)\tanh^{2}(\zeta/2)}}\biggr)\nonumber\\
\ \ \ \ \ \ \ \ \ \ \ \ \ \ \ \ \ \ \ \ \ \ \ \ \ \ \ \ \ \ \ \ \ \ \ \ \ \ \ \ \ \ \ \ \mbox{for}\ \biggl(\tan(x)<0\biggr)\label{eq:unstable_minus}.
\end{eqnarray}
In this case, another Bethe quantum number $J_{2}$ is equal to $J_{1}$: $J_{2}=J_{1}$. 
When the inequality $\frac{N-1}{2}<\frac{N}{\pi}\tan^{-1}\biggl(\sqrt{\frac{N-(1+\tanh^{2}(\zeta/2))}{1-(N-1)\tanh^{2}(\zeta/2)}}\biggr)$ is satisfied, the extra two-string solution appear.
The Bethe quantum numbers corresponding to the extra two-string solution are $(\frac{N-1}{2},\frac{N-1}{2})$ and $(-\frac{N-1}{2},-\frac{N-1}{2})$

\item The singular solution

When $N=4n$ with an integer $n$, the Bethe quantum numbers are
\begin{eqnarray}
(J_{1}, J_{2})=\Bigl(\frac{N}{4}-\frac{1}{2},  \frac{N}{4}+\frac{1}{2}\Bigr).
\end{eqnarray}
Similarly, when $N=4n+2$ with an integer $n$, the Bethe quantum numbers are
\begin{eqnarray}
(J_{1}, J_{2})=\Bigl(\frac{N}{4}, \frac{N}{4}\Bigr).
\end{eqnarray}
\end{enumerate}

Summarizing the above, we obtain the Bethe quantum number for the complex solution as shown in the Figure \ref{fig:complex_bqe_list}.

\begin{figure}[htbp]
\begin{center}
 \includegraphics[clip,width=15cm]{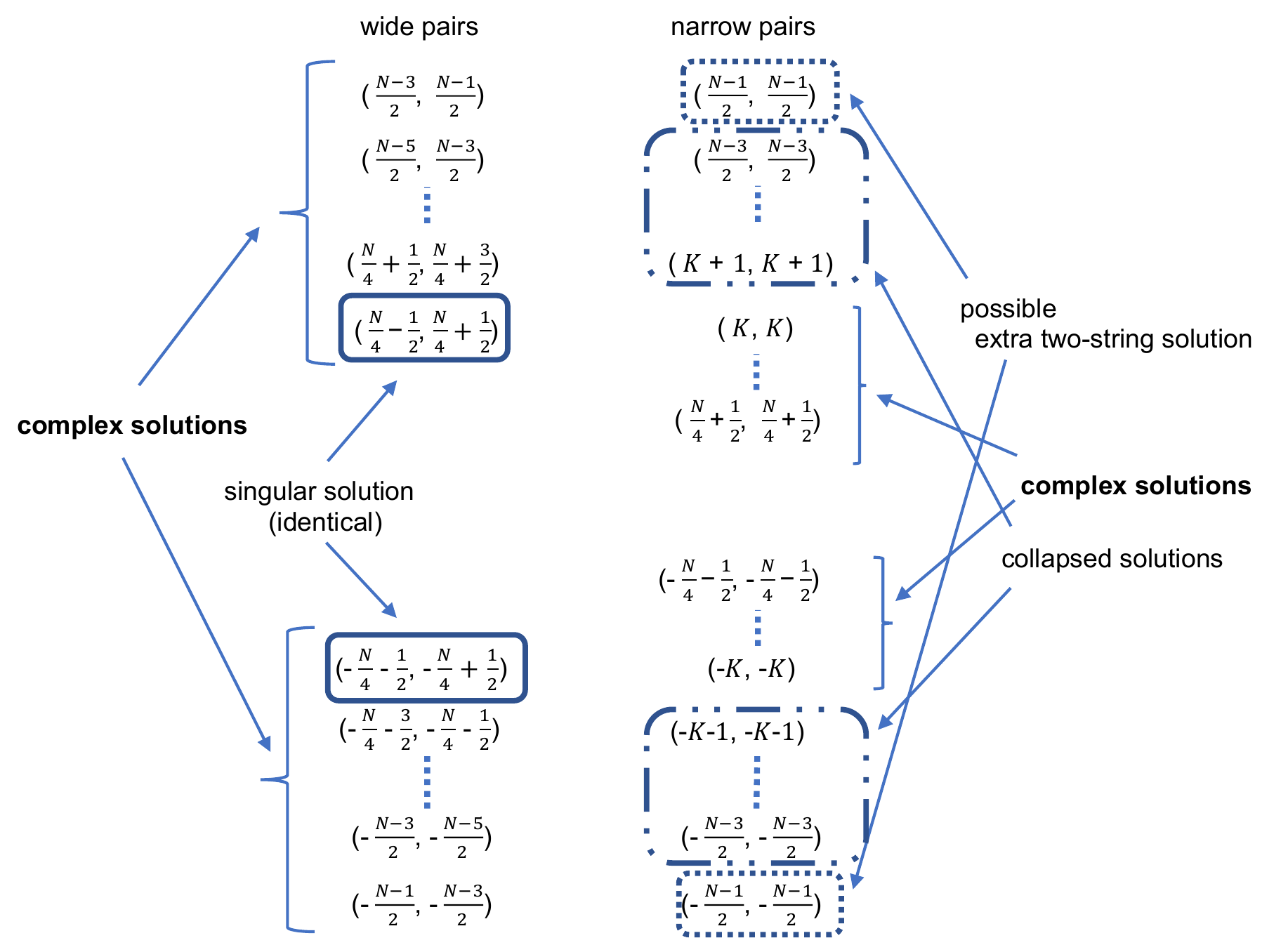}
\end{center}
 \caption{
 Illustrates Bethe quantum numbers for the complex solutions, where $N=4n$ with an integer $n$.
If the parameters $\zeta, N$ are in the unstable regime, $K$ is defined by
 $K\equiv\biggl[\frac{N}{\pi}\tan^{-1}\Bigl(\sqrt{\frac{N-(1+\tanh^{2}(\zeta/2))}{1-(N-1)\tanh^{2}((\zeta/2)}}\Bigr)\biggr]_{Gauss}-\frac{1}{2}$.
On the other hand, if the parameters $\zeta, N$ are in stable regime , $K$ is defined by $K\equiv\frac{N-1}{2}$.
 We remark the two singular solutions exist but they are same valued solutions. 
 In addition, the number of the narrow pair depend on the value of $K$.
When $K=\frac{N-1}{2}$, the extra two-string solutions emerge.  
When $K<\frac{N-3}{2}$, the solutions corresponding to $(\frac{N-3}{2},\frac{N-3}{2}) \cdots (K+1, K+1)$ and $(-K-1, -K-1)\cdots(-\frac{N-3}{2},-\frac{N-3}{2})$ are collapsed.
 }
 \label{fig:complex_bqe_list} 
\end{figure}

\begin{figure}[htbp]
\begin{center}
 \includegraphics[width=15cm]{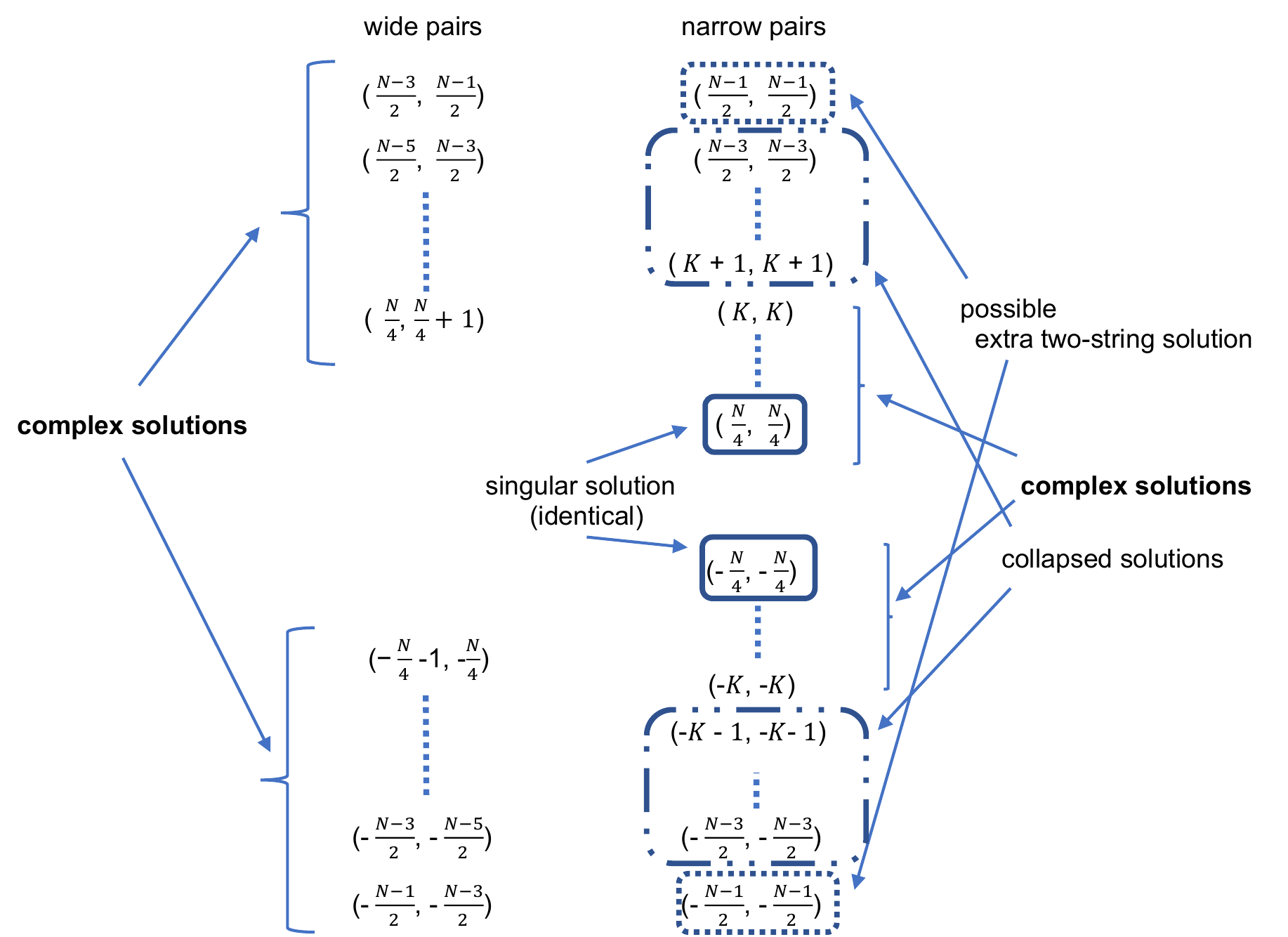}
\end{center}
 \caption{
Illustrates the Bethe quantum numbers for complex solutions, where $N=4n+2$ with an integer $n$.
 }
 \label{fig:complex_bqe_list_4n+2} 
\end{figure}

\item The finite Bethe solution for the Bethe ansatz equation in the XXX limit\\

In this case, we divide the solutions of the Bethe ansatz equation into those corresponding to m=0 and those corresponding to m>0. We recall that $m$ is the difference between two Bethe quantum numbers (i.e. $m=J_{1}-J_{2}$).
We define the difference of the Bethe quantum numbers $J_{1}$ and $J_{2}$ as $m$: $m=J_{2}-J_{1}$.
We divide the Bethe quantum numbers corresponding to the difference of the Bethe quantum numbers $J_{1}$ and $J_{2}$.

\begin{enumerate}[(i)]\label{enum:normal_real_solution}
\item The difference of Bethe quantum numbers $m=1,\cdots,N-1$\label{enum:normal_real_solution}\\
From (\ref{eq:finite_bqn_set}) in section 5, the set of the Bethe quantum numbers $J_{1}$ and $J_{2}$ satisfy the following conditions:
\begin{eqnarray}
-\frac{N-1}{2}<J_{1}<J_{2}<\frac{N-1}{2}
\end{eqnarray}
They are the conditions of the Bethe quantum numbers for the standard one-string solutions.
In this case, another Bethe quantum number is equal to $J_{1}+m$: $J_{2}=J_{1}+m$.
\item The difference of Bethe quantum numbers $m=0$\label{enum:m=0_real_solution}\\
In the condition $J_{1}=J_{2}$, we need to consider the extra or collapsed two-string solutions. 
From (\ref{eq:extra_two_string_condition}), (\ref{eq:collapse_condition}), the Bethe quantum number $J_{1}$ satisfy
\begin{eqnarray}
\frac{N}{\pi}\tan^{-1}&&\biggl(\sqrt{\frac{N-(1+\tanh^{2}(\zeta/2))}{1-(N-1)\tanh^{2}(\zeta/2)}}\biggr)\nonumber\\
&&\leq J_{1}<\frac{N}{2}-\frac{1}{2}\ \ \mbox{for}\ \Bigl(\tan(x)>0\biggr),
\end{eqnarray}  

\begin{eqnarray}
-\frac{N}{2}+\frac{1}{2}<J_{1}\nonumber\\
\leq-\frac{N}{\pi}\tan^{-1}\biggl(\sqrt{\frac{N-(1+\tanh^{2}(\zeta/2))}{1-(N-1)\tanh^{2}(\zeta/2)}}\biggr)\ \ \mbox{for}\ \Bigl(\tan(x)<0\biggr)\nonumber\\
\end{eqnarray}
In this case another Bethe quantum number $J_{2}$ is equal to $J_{1}$: $J_{2}=J_{1}$.
corresponding to collapsed solution
\end{enumerate}
\item The infinite Bethe solution for the Bethe ansatz equation\label{enum:enum:inf_solution}\\

In this case, the Bethe quantum numbers is given by the section \ref{sec:divergence}.
These Bethe quantum numbers include that corresponding to the extra two-string solutions. 
From (\ref{eq:bqn_infinite_positive}) and (\ref{eq:bqn_infinite_negative}), when the site number $N$ is even, the sets of Bethe quantum numbers are 
\begin{eqnarray}
(J_{1}, J_{2})=\biggl(\frac{N-1}{2},\ \frac{1}{2}\biggr),\ \biggl(\frac{N-1}{2},\ \frac{3}{2}\biggr)\cdots\nonumber\\
\ \ \ \ \ \ \ \ \ \ \ \ \ \ \ \ \ \ \ \ \ \cdots,\biggl(\frac{N-1}{2},\ \frac{N-1}{2}\biggr)\ \mbox{for}\ \Bigl(\tan(x)>0\biggr)\\
(J_{1}, J_{2}) =\biggl(-\frac{N-1}{2},-\frac{1}{2}\biggr),\biggl(-\frac{N-1}{2},-\frac{3}{2}\biggr)\cdots\nonumber\\
\ \ \ \ \ \ \ \ \ \ \ \ \ \ \ \ \ \cdots,\biggl(-\frac{N-1}{2},-\frac{N-1}{2}\biggr)\ \mbox{for}\ \Bigl(\tan(x)<0\biggr).
\end{eqnarray}

When the inequality $\frac{N-1}{2}<\frac{N}{\pi}\tan^{-1}\biggl(\sqrt{\frac{N-(1+\tanh^{2}(\zeta/2))}{1-(N-1)\tanh^{2}(\zeta/2)}}\biggr)$ is satisfied, the extra two-string solution appears.
The Bethe quantum number $\bigl(\frac{N-1}{2},\ \frac{N-1}{2}\bigr)$ and $\bigl(-\frac{N-1}{2},-\frac{N-1}{2}\bigr)$ correspond to the extra two-string solutions.
Thus, when the extra two-string solution appears, the Bethe solution corresponding to the Bethe quantum number $\bigl(\frac{N-1}{2},\ \frac{N-1}{2}\bigr)$ and $\bigl(-\frac{N-1}{2},-\frac{N-1}{2}\bigr)$ become the complex solutions.
It corresponds to \ref{enum:unstable}.

On the other hand, when the site number $N$ is odd, the sets of the Bethe quantum numbers are 
\begin{eqnarray}
(J_{1}, J_{2})=\biggl(\frac{N-1}{2},\ 0\biggr),\ \biggl(\frac{N-1}{2},\ 1\biggr)\cdots\nonumber\\
\ \ \ \ \ \ \ \ \ \ \ \ \ \ \ \ \ \ \cdots,\biggl(\frac{N-1}{2},\ \frac{N-1}{2}\biggr)\ \mbox{for}\ \Bigl(\tan(x)>0\biggr)\\
(J_{1}, J_{2}) =\biggl(-\frac{N-1}{2},-1\biggr),\biggl(-\frac{N-1}{2},-2\biggr)\cdots\nonumber\\
\ \ \ \ \ \ \ \ \ \ \ \ \ \ \cdots,\biggl(-\frac{N-1}{2},-\frac{N-1}{2}\biggr)\ \mbox{for}\ \Bigl(\tan(x)<0\biggr).
\end{eqnarray}
The Bethe quantum number $\bigl(\frac{N-1}{2},\ \frac{N-1}{2}\bigr)$ and $\bigl(-\frac{N-1}{2},-\frac{N-1}{2}\bigr)$ correspond to the extra two-string solutions.
\end{enumerate}
Summarizing the above, we obtain the Bethe quantum number for the real solution as shown in the figure \ref{fig:bqe_list}. 

\subsection{The example of the Bethe quantum number list}
In this subsection, we show the example of the list for the Bethe quantum numbers corresponding to the real solution in the case of $N=8$ and $\zeta=0.6$

\begin{enumerate}[(I)]
\item Complex solution

\begin{enumerate}[(i)]

\item Narrow pair

In this parameter(i. e. $N=8$ and $\zeta=0.6$), we estimate
\begin{eqnarray}
\tanh^{2}(\zeta/2)-\frac{1}{N-1}=-0.0579941<0.
\end{eqnarray}
Thus, we consider an unstable regime.
From (\ref{eq:unstable_plus}) and (\ref{eq:unstable_minus}), the Bethe quantum numbers $J_{1}$ for narrow pairs are given by
\begin{eqnarray}
\frac{N}{4}=2\leq J_{1} \leq \frac{N}{\pi}\tan^{-1}\biggl(\sqrt{\frac{N-(1+t^{2})}{1-(N-1)t^{2}}}\biggr)=3.39467\nonumber\\
\\
-\frac{N}{\pi}\tan^{-1}\biggl(\sqrt{\frac{N-(1+t^{2})}{1-(N-1)t^{2}}}\biggr)=-3.39467<J_{1}\leq-\frac{N}{4}=-2\nonumber\\
\end{eqnarray}
We have the Bethe quantum numbers $(5/2, 5/2)$ and $(-5/2, -5/2)$

\item Wide pair

In this case, 
we obtain the Bethe quantum number from (\ref{eq:wide_pair_plus}) and (\ref{eq:wide_pair_minus}).
We estimate
\begin{eqnarray}
\frac{3}{2}<J_{1}<\frac{7}{2},\\
-\frac{9}{2}<J_{1}<-\frac{5}{2}.
\end{eqnarray}
Thus, we have $(5/2, 7/2)$, $(-7/2, -5/2)$

\item Singular solution\\
We have 
\begin{eqnarray}
(J_{1}, J_{2})=\Bigl(\frac{3}{2}, \frac{5}{2}\Bigr).
\end{eqnarray}
\end{enumerate}

Summarizing the above, we obtain the Bethe qunatum number for the complex solution as shown in the Figure \ref{fig:example_complex_bqe_list}.

\begin{figure}[htbp]
\begin{center}
 \includegraphics[clip,width=8cm]{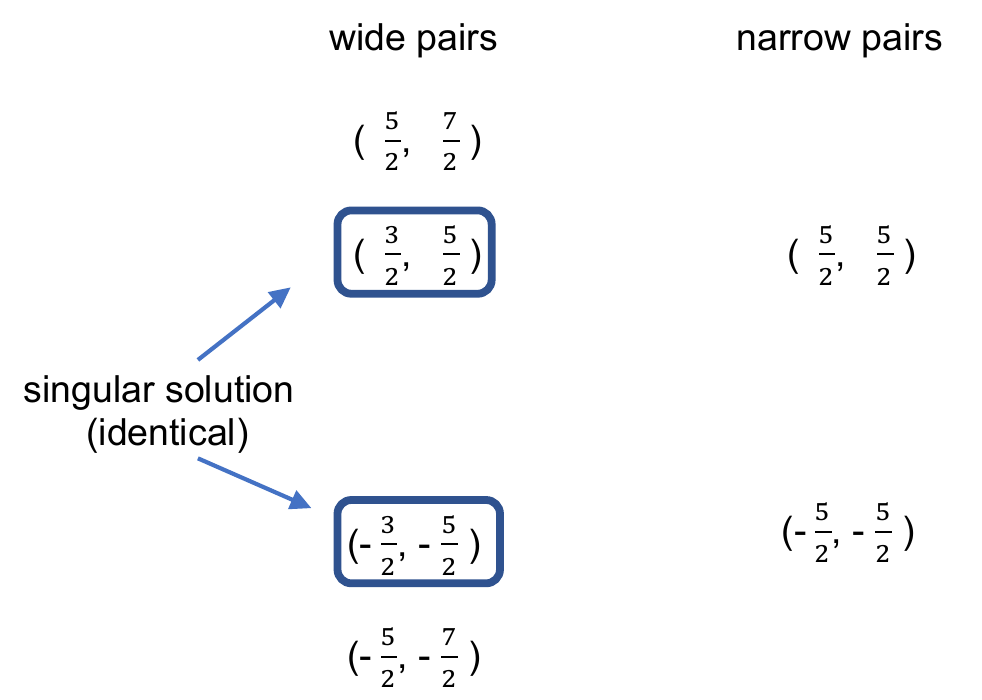}
\end{center}
\caption{Bethe quantum numbers for complex solutions for $N=8$ and $\zeta=0.6$.
We remark that the two pairs of Bethe quantum numbers for the singular solution exist but they are equivalent. 
}
 \label{fig:example_complex_bqe_list} 
\end{figure}

\item Real solution
\begin{enumerate}[(i)]
\item The standard real solutions $\Bigl(i.e.(-\frac{N-3}{2}\leq)J_{1}<J_{2}(\leq\frac{N-3}{2})\Bigr)$\\
The Bethe quantum numbers for the real solution satisfying 
\begin{eqnarray}
\biggl(-\frac{N-3}{2}\leq)J_{1}<J_{2}(\leq\frac{N-3}{2}\biggr)
\end{eqnarray}
are $(-5/2, -3/2)$, $(-5/2, -1/2)$, $(-5/2, 1/2)$, $(-5/2, 3/2)$, $(-5/2, 5/2)$, $(-3/2, -1/2)$, $(-3/2, 1/2)$, $(-3/2, 3/2)$, $(-3/2, 5/2)$, $(-1/2, 1/2)$, $(-1/2, 3/2)$, $(-1/2, 5/2)$, $(1/2, 3/2)$, $(1/2, 5/2)$, $(3/2, 5/2)$.

\item Infinite Bethe solution($J_{1}\neq J_{2}$)\\
The Bethe quantum number for the infinite Bethe solutions satisfying $J_{1}\neq J_{2}$ is 
$(-1/2, -7/2)$, $(-3/2, -7/2)$, $(-5/2, -7/2)$, $(1/2, 7/2)$, $(3/2, 7/2)$, $(5/2, 7/2)$.

\item Infinite Bethe solution($J_{1}=J_{2}$, $J_{1}=\frac{N-1}{2} \mbox{or} -\frac{N-1}{2}$)\\

In this parameter(i. e. $N=8$ and $\zeta=0.6$), we estimate
\begin{eqnarray}
\frac{N}{\pi}\tan^{-1}\biggl(\sqrt{\frac{N-(1+t^{2})}{1-(N-1)t^{2}}}\biggr)=3.39467\cdots<\frac{N-1}{2}=\frac{7}{2}.\nonumber\\
\end{eqnarray}
Thus, the extra two-string solution does not exist.
There exist the Bethe quantum number $(\frac{N-1}{2},\frac{N-1}{2})$ and $(-\frac{N-1}{2},-\frac{N-1}{2})$ as real solution.
Thus,  the Bethe quantum number corresponding to the infinite Bethe solution ($J_{1}=J_{2}$) is $(7/2, 7/2)$, $(-7/2, -7/2)$.
\end{enumerate}

Summarizing the above, we obtain the Bethe quantum number for the real solution as shown in Figure \ref{fig:real_bqe_example_list}. 

\begin{figure}[htbp]
\begin{center}
 \includegraphics[clip,width=13cm]{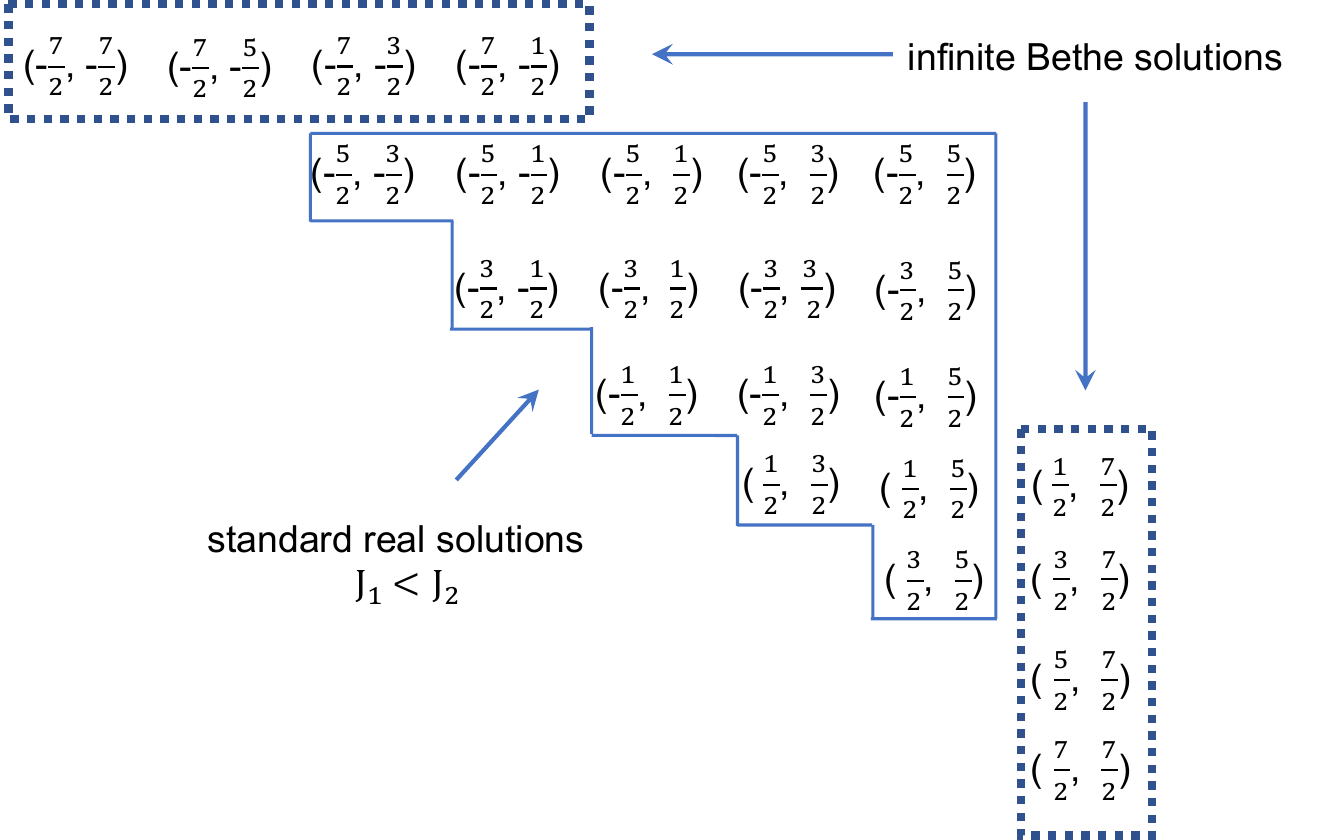}
\end{center}
 \caption{
Illustrate the Bethe quantum number for the real solutions for $N=8$ and $\zeta=0.6$.
}
 \label{fig:real_bqe_example_list} 
\end{figure}

\end{enumerate}


\section{New method to derive a solution of the Bethe ansatz equations in the two down-spin sector}\label{sec:get_solution}

In this section, we show how to get numerically the exact solution of the Bethe ansatz equations from the Bethe quantum numbers.

\subsection{The complex solution: Counting function method} 

In the complex solution, there exist two methods to obtain the solution of the Bethe ansatz equation.
The first method is an iterative approach.
The Bethe quantum number is essential for obtaining exact solutions with iterative method.
The second is a counting function approach.
We introduce the counting function for the complex solution approach. 


This method is proposed in the \cite{ISD}.
The counting function for the complex solution is defined by
\begin{eqnarray}
Z_{1}(\delta(w), x(w), \zeta)&&:=\frac{1}{2\pi}\tan^{-1}\Bigl(\frac{a}{1-b}\Bigr)+\frac{1}{2\pi}\tan^{-1}\Bigl(\frac{a}{1+b}\Bigr)\nonumber\\
&&+\frac{1}{2}\Bigl(H(b-1)+2H(1-b)H(-a)-\frac{1}{N}H(\delta)\Bigr)\label{eq:complex_counting_func}
\end{eqnarray}
where a and b is given by
\begin{eqnarray}
a=&&\frac{\tan x(1-w^{2}t^{2})}{t(1+(\tan^{2}x)w^{2}t^{2})}\\
b=&&\frac{(1+\tan^{2}{x})w}{(1+(\tan^{2}x))w^{2 t^{2}}}
\end{eqnarray}
and $w$ and $t$ is defined by
\begin{eqnarray}
t=&&\tanh(\zeta/2)\\
w=&&\frac{\tanh(\zeta/2+\delta)}{\tanh(\zeta/2)}.
\end{eqnarray}
The string center $x$ is given by
\begin{eqnarray}
\tan^{2}x=\frac{1}{2A(w)}(-B(w)-\sqrt{B(w)^{2}-4A(w)C(w)})
\end{eqnarray}
where $A(w)$, $B(w)$, $C(w)$ is defined by
\begin{eqnarray}
A(w)=&&w^{2}(1+wt^{2})^{2}\biggl\{\Bigl(\frac{-(1-w)(1-wt^{2})}{1+w^{2}t^{2}}\Bigr)^{2}\biggr\}^{\frac{1}{N}}\nonumber\\
&&-w^{2}(1-wt^{2})^{2}\biggl\{\Bigl(\frac{(1+w)(1+wt^{2})}{1+w^{2}t^{2}}\Bigr)^{2}\biggr\}^{\frac{1}{N}}\\
B(w)=&&\Bigl\{\frac{(1-w^{2}t^{2})^{2}}{t^{2}}+2w(1+w)(1+wt^{2})\Bigr\}
\biggl\{\Bigl(\frac{-(1-w)(1-wt^{2})}{1+w^{2}t^{2}}\Bigr)^{2}\biggr\}^{\frac{1}{N}}\nonumber\\
&&-\Bigl\{\frac{(1-w^{2}t^{2})^{2}}{t^{2}}-2w(1-w)(1-wt^{2})\Bigr\}
\biggl\{\Bigl(\frac{(1+w)(1+wt^{2})}{1+w^{2}t^{2}}\Bigr)^{2}\biggr\}^{\frac{1}{N}}\nonumber\\
\\
C(w)=&&(1+w)^{2}\biggl\{\Bigl(\frac{-(1-w)(1-wt^{2})}{1+w^{2}t^{2}}\Bigr)^{2}\biggr\}^{\frac{1}{N}}\nonumber\\
&&-(1-w)^{2}\biggl\{\Bigl(\frac{(1+w)(1+wt^{2})}{1+w^{2}t^{2}}\Bigr)^{2}\biggr\}^{\frac{1}{N}}.
\end{eqnarray}

We remark that the counting function is a function of only $w$.
Using the counting function (\ref{eq:complex_counting_func}), the Bethe ansatz equation is expressed by the counting function as follows.
\begin{eqnarray}
Z_{1}(\delta(w), x(w), \zeta)=\frac{J_{1}}{N}.\label{eq:BAE_complex_counting}
\end{eqnarray}
We need to solve this type of Bethe ansatz equation (\ref{eq:BAE_complex_counting}).

We consider the three cases: (\si) If $w<1$ holds, then the counting function is continuous and decreasing monotone on the domain of definition for itself.
We remark that this case corresponds to a narrow pair.
On the other hand, $w$ is more than $1$;(\sii)If $w>1$ holds, then the counting function is continuous and increasing monotone on the domain of definition for itself.
We remark that this case corresponds to a wide pair.
Thus, we solve this equation numerically without the singular solution.
In addition, (\siii) the singular solution is expressed as follows.
\begin{eqnarray}
\lambda_{1,2}=\pm\frac{1}{2}\zeta i
\end{eqnarray}

\subsection{Real solutions for $J_{1}=J_{2}$: Counting function method}

There are two ways to get the real solution for $J_{1}=J_{2}$.
One is the iteration method, the other is the counting function approach.
We introduce the counting function approach for the real solution.


The counting function for the real solution is defined by (\ref{eq:real_part_right_second_term}).
The Bethe ansatz equation expressed using the counting function for the real solution is (\ref{eq:BAE_counting_function_real}).
When $J_{1}-J_{2}=0$ holds, the counting function is shown in Fig\ref{fig:counting_function_for_the_real_and_the_complex_solution}.
Thus, we can solve this equation with some methods such as the bisection method and the Newton method in this case.

\subsection{Real solution for $J_{1}\neq J_{2}$: Height function method}

There are two methods to obtain the real solution for $J_{1}\neq J_{2}$.
The first method is an iterative approach.
The second is a counting function approach.
We introduce the height function approach. 


The height function is defined by the equation(\ref{eq:def_search_func}).
The Bethe ansatz equations expressed in terms of the height function are (\ref{eq:lambda2_with_lambda1}) and (\ref{eq:BAE_height_function}).
The height function is continuous and decreasing monotone except for the specific point $\lambda_{1}=K^{J_{1}}$, which is discussed in the section \ref{sec:monotonicity_of_the_height_function}.
Thus, we can solve this equation by the bisection method or the Newton method by choosing an initial value near $\lambda^{J_{1}}$ and $\lambda^{J_{1}+1}$.

\ack

The present research was partially supported by the Grant-in-Aid for Scientific Research (C) (21K03398).

\appendix

\section{The proof of $\displaystyle \Bigl(-\frac{\pi}{2}<\Bigr)\bar{\lambda}_{1}\zeta-\bar{\lambda}_{2}\zeta<\frac{\pi}{2}$ in $\displaystyle J_{1}=\frac{N-1}{2}$}\label{sec:app_ineq}

In this section, we show that when the quantum number $J_{1}=\frac{N-1}{2}$, $\bar{\lambda}_{1},\bar{\lambda}_{2}$ satisfy the relation$(-\frac{\pi}{2}<)\bar{\lambda}_{1}\zeta-\bar{\lambda}_{2}\zeta<\frac{\pi}{2}$ Using a contradiction.
We assume that the inequality $\frac{\pi}{2}<\bar{\lambda}_{1}\zeta-\bar{\lambda}_{2}\zeta$. We consider the first Bethe ansatz equation (\ref{eq:BAE11}) in this case.

\begin{eqnarray}
2\tan^{-1}\biggl(\frac{\tan(\bar{\lambda_{1}\zeta})}{\tanh(\frac{\zeta}{2})}\biggr)=\frac{2\pi}{N}\frac{N-1}{2}+\frac{2}{N}\tan^{-1}\biggl(\frac{\tan(\bar{\lambda}_{1}\zeta-\bar{\lambda}_{2}\zeta-\pi)}{\tanh(\zeta)}\biggr)+\frac{2\pi}{N}.\nonumber\\
\end{eqnarray}
It is equivalent to
\begin{eqnarray}
\tan^{-1}\biggl(\frac{\tan(\bar{\lambda}_{1}\zeta-\bar{\lambda}_{2}\zeta-\pi)}{\tanh(\zeta)}\biggr)=N\tan^{-1}\biggl(\frac{\tan(\bar{\lambda_{1}\zeta})}{\tanh(\frac{\zeta}{2})}\biggr)-\pi\frac{N+1}{2}\nonumber\\
\end{eqnarray}
and
\begin{eqnarray}
-\frac{\pi}{2}<N\tan^{-1}\biggl(\frac{\tan(\bar{\lambda}_{1}\zeta)}{\tanh(\frac{\zeta}{2})}\biggr)-\pi\frac{N+1}{2}<\frac{\pi}{2}.
\end{eqnarray}
Thus
\begin{eqnarray}
\tan^{-1}\biggl(\frac{\tan(\bar{\lambda}_{1}\zeta-\bar{\lambda}_{2}\zeta-\pi)}{\tanh(\zeta)}\biggr)=N\tan^{-1}\biggl(\frac{\tan(\bar{\lambda_{1}\zeta})}{\tanh(\frac{\zeta}{2})}\biggr)-\pi\frac{N+1}{2}\nonumber\\
\end{eqnarray}
and
\begin{eqnarray}
\frac{\pi}{2}=\frac{\pi}{2N}+\frac{\pi}{2}\frac{N+1}{N}<\tan^{-1}\biggl(\frac{\bar{\lambda}_{1}\zeta}{\tanh(\frac{\zeta}{2})}\biggr)<\frac{\pi}{2N}+\frac{\pi}{2}\frac{N+1}{N}=\frac{N+2}{2N}\pi.\nonumber\\\label{eq:app_contradiction_ineq}
\end{eqnarray}

A contradiction is obtained by (\ref{eq:app_contradiction_ineq})
$-\frac{\pi}{2}<\tan^{-1}(x)<\frac{\pi}{2}$.
Thus, we obtain $\bar{\lambda}_{1}\zeta-\bar{\lambda}_{2}\zeta<\frac{\pi}{2}$.
On the other hand, from $\bar{\lambda}_{1}>0$ it is clear that $-\frac{\pi}{2}<\bar{\lambda}_{1}\zeta-\bar{\lambda}_{2}\zeta$.
Therefore, we obtain $-\frac{\pi}{2}<\bar{\lambda}_{1}\zeta-\bar{\lambda}_{2}\zeta<\frac{\pi}{2}$.


\section{
The case where the first solution $\lambda_{1}$ of the Bethe ansatz equation is negative}\label{sec:negative_lambda1}
We consider the case where the first rapidity $\lambda_{1}$ is negative.
We define $(\tilde{\lambda}_{1},\tilde{\lambda}_{2})$ as $(\tilde{\lambda}_{1},\tilde{\lambda}_{2})\equiv(-\lambda_{1},-\lambda_{2})$ and $(\tilde{J}_{1}, \tilde{J}_{2})$ is the Bethe quantum number corresponding to $(\tilde{\lambda}_{1},\tilde{\lambda}_{2})$.
We substitute $(\tilde{\lambda}_{1},\tilde{\lambda}_{2})$ and $\tilde{J}_{1}$ into the first Bethe ansatz equation (\ref{eq:BAE11}) as follows.
\begin{eqnarray}
     \frac{2\pi}{N}\tilde{J}_{1}&=&-2\tan^{-1}\biggl(\frac{\tan\tilde{\lambda}_{1}}{\tanh{\zeta/2}}\biggr)+\frac{2}{N}\tan^{-1}\biggl(\frac{\tan(\tilde{\lambda}_{1}-\tilde{\lambda}_{2})}{\tanh\zeta}\biggr)+\frac{2\pi}{N}\biggl[\frac{2(\tilde{\lambda}_{1}-\tilde{\lambda}_{2})+\pi}{2\pi}\biggr]_{Gauss}\nonumber\\
     &=&2\tan^{-1}\biggl(\frac{\tan\lambda_{1}}{\tanh{\zeta/2}}\biggr)-\frac{2}{N}\tan^{-1}\biggl(\frac{\tan(\lambda_{1}-\lambda_{2})}{\tanh\zeta}\biggr)-\frac{2\pi}{N}\biggl[\frac{2(\lambda_{1}-\lambda_{2})+\pi}{2\pi}\biggr]_{Gauss}\nonumber\\
     &=&-\frac{2\pi}{N}J_{1}
\end{eqnarray}
The same argument can be applied to the Bethe equation for the Bethe qunatum number $\tilde{J}_{2}$ (\ref{eq:BAE22}).
Thus, if we have the Bethe quantum number $(J_{1}, J_{2})$ and the solution of Bethe ansatz equation $(\lambda_{1}, \lambda_{2})$ corresponding to them, $(-\lambda_{1}, -\lambda_{2})$ is solution of Bethe ansatz equation corresponding to Bethe quantum number $(-J_{1}, -J_{2})$.

\section{Difference of the two Bethe quantum numbers $J_{1}$,  $J_{2}$}\label{sec:massive_real_quantum}
In the section \ref{sec:intro_counting}, we derived the counting function corresponding to the first Bethe ansatz equation (\ref{eq:BAE11}). In this appendix, we show the relation between the Bethe quantum numbers $J_{1}$ and $J_{2}$.

\begin{eqnarray}
\frac{2\pi}{N}(J_{1}-J_{2})&=&\frac{1}{2i}\log\biggl\{\frac{(\tanh(\frac{\zeta}{2})+i\tan(x-\frac{1}{2}\gamma\zeta))(\tanh(\frac{\zeta}{2})-i\tan(x+\frac{1}{2}\gamma\zeta))}{(\tanh(\frac{\zeta}{2})-i\tan(x-\frac{1}{2}\gamma\zeta))(\tanh(\frac{\zeta}{2})+i\tan(x+\frac{1}{2}\gamma\zeta))}\biggr\}\nonumber\\
&-&\frac{1}{2iN}\biggl\{\log\biggl(\frac{\tanh(\zeta)+i\tan(-\gamma\zeta)}{\tanh(\zeta)-i\tan(-\gamma\zeta)}\biggr)^{2}\biggr\}\nonumber\\
&-&\frac{1}{2i}\log\biggl\{\frac{(\tanh(\frac{\zeta}{2})+i\tan(x+\frac{1}{2}\gamma\zeta))(\tanh(\frac{\zeta}{2})-i\tan(x-\frac{1}{2}\gamma\zeta))}{(\tanh(\frac{\zeta}{2})-i\tan(x+\frac{1}{2}\gamma\zeta))(\tanh(\frac{\zeta}{2})+i\tan(x-\frac{1}{2}\gamma\zeta))}\biggr\}\nonumber\\
&+&\frac{1}{2iN}\biggl\{\log\biggl(\frac{\tanh(\zeta)+i\tan(\gamma\zeta)}{\tanh(\zeta)-i\tan(\gamma\zeta)}\biggr)^{2}\biggr\}\nonumber\\
&=&\frac{1}{2i}\log\biggl\{\frac{(\tanh(\frac{\zeta}{2})+i\tan(x-\frac{1}{2}\gamma\zeta))(\tanh(\frac{\zeta}{2})-i\tan(x+\frac{1}{2}\gamma\zeta))}{(\tanh(\frac{\zeta}{2})-i\tan(x-\frac{1}{2}\gamma\zeta))(\tanh(\frac{\zeta}{2}+i\tan(x+\frac{1}{2}\gamma\zeta))}\nonumber\\
&&\ \ \ \ \ \ \ \ \ \ \ \ \ \ \times\frac{(\tanh(\frac{\zeta}{2})-i\tan(x+\frac{1}{2}\gamma\zeta))(\tanh(\frac{\zeta}{2})+i\tan(x-\frac{1}{2}\gamma\zeta))}{(\tanh(\frac{\zeta}{2})+i\tan(x+\frac{1}{2}\gamma\zeta))(\tanh(\frac{\zeta}{2})-i\tan(x-\frac{1}{2}\gamma\zeta))}\biggr\}\nonumber\\
&+&\frac{1}{iN}\log\biggl\{\biggl(\frac{\tanh(\zeta)+i\tan(\gamma\zeta)}{\tanh(\zeta)-i\tan(\gamma\zeta)}\biggr)^{2}\biggr\}\nonumber\\
&=&\frac{1}{i}\log\biggl\{\frac{(\tanh(\frac{\zeta}{2})+i\tan(x-\frac{1}{2}\gamma\zeta))(\tanh(\frac{\zeta}{2})-i\tan(x+\frac{1}{2}\gamma\zeta))}{(\tanh(\frac{\zeta}{2})-i\tan(x-\frac{1}{2}\gamma\zeta))(\tanh(\frac{\zeta}{2})+i\tan(x+\frac{1}{2}\gamma\zeta))}\biggr\}\nonumber\\
&+&\frac{1}{iN}\log\biggl\{\biggl(\frac{\tanh(\zeta)+i\tan(\gamma\zeta)}{\tanh\zeta-i\tan(\gamma\zeta)}\biggr)^{2}\biggr\}\nonumber\\
&=&\frac{1}{i}\log\exp\biggl(\frac{2\pi in}{N}\biggr)\nonumber\\
&=&\frac{2\pi}{N}n
\end{eqnarray}
In the forth step we use the following equality
\begin{eqnarray}
\frac{1}{iN}\log\biggl\{\biggl(\frac{\tanh(\zeta)+i\tan(\gamma\zeta)}{\tanh(\zeta)-i\tan(\gamma\zeta)}\biggr)^{2}\biggr\}=\frac{1}{i}\log\biggl\{\exp\Bigl(\frac{2\pi in}{N}\Bigr)\biggl(\frac{\tanh(\zeta)+i\tan(\gamma\zeta)}{\tanh\zeta-i\tan(\gamma\zeta)}\biggr)^{\frac{2}{N}}\biggr\}.\nonumber\\
\end{eqnarray}
Thus, we obtain
\begin{eqnarray}
\therefore J_{1}-J_{2}=n\ \ \ \ (n=0,1,2,\cdots,N-1).
\end{eqnarray}
There exist each cases $n=0,-1,-2,\cdots,-(N-1)$. 
Therefore we derived the correspondence between the difference of the quantum Bethe numbers $J_{1}$, $J_{2}$ and $n=0,-1,-2,\cdots,-(N-1)$.  
We should note that we cannot obtain all Bethe quantum numbers except for $J_{1}=J_{2}$ case.

\end{document}